\documentclass[]{mn2e}
\usepackage{psfig, epsf, epsfig}
\title[Formation of GCSs]{On the Origin of Mass--Metallicity Relations, 
Blue Tilts, and Scaling Relations for Metal-poor Globular Cluster Systems}
\author[K. Bekki, H. Yahagi, and D. A. Forbes]
       {Kenji Bekki${}^1$\thanks{E-mail: bekki@bat.phys.unsw.edu.au}
        Hideki Yahagi${}^2$\thanks{E-mail: hyahagi@astron.s.u-tokyo.ac.jp},
        and Duncan A. Forbes${}^3$\thanks{E-mail: dforbes@astro.swin.edu.au} \\ 
        ${}^1$School of Physics, University of New South Wales,
              Sydney 2052, NSW, Australia\\
        ${}^2$Department of Astronomy, University of Tokyo,
              7-3-1 Hongo, Bunkyo ward, Tokyo 113-0033, Japan\\
        ${}^3$Centre for Astrophysics \& Supercomputing,
Swinburne University of Technology,
Hawthorn, VIC 3122, Australia}

\begin{document}

\date{Accepted, Received 2005 May 13; in original form }

\pagerange{\pageref{firstpage}--\pageref{lastpage}} \pubyear{2005}

\maketitle

\label{firstpage}

\begin{abstract}

We investigate formation processes and physical properties of
globular cluster systems (GCSs) in galaxies based on
high-resolution cosmological simulations with globular
clusters.
% forming at high redshifts ($z>6$).
We focus on metal-poor clusters
(MPCs) and correlations with their host galaxies by assuming
that MPC formation is truncated at a high redshift ($z_{\rm trun}
\ge 6$). 
We find that the correlation between mean metallicities
($Z_{\rm gc}$) of MPCs and their host galaxy luminosities
($L$) flattens from $z=z_{\rm trun}$ to $z=0$.
We also find that the observed relation
($Z_{\rm gc} \propto L^{0.15}$) in MPCs
can be reproduced well in the models with
$Z_{\rm gc} \sim L^{0.5}$ at   $z=z_{\rm trun}$ when 
$z_{\rm trun} \sim 10$,
if mass-to-light-ratios are assumed to be constant 
at $z=z_{\rm trun}$. 
A flatter $L-Z_{\rm gc}$ at   $z=z_{\rm trun}$
is found to be required to explain the observed relation
for constant mass-to-light-ratio  models with lower $z=z_{\rm trun}$.
However, better agreement with the 
observed relation 
%$Z_{\rm gc} \propto L^{0.15}$ 
is found for models with
different mass-to-light-ratios between $z=z_{\rm trun}$
and $z=0$.
% for a given $z_{\rm trun}$.
It is also found that the observed color-magnitude
relation of luminous MPCs (i.e., ``blue tilts'')
may only have a small contribution from the stripped stellar nuclei
of dwarf galaxies, which have nuclei masses that correlate with
their total mass at $z=z_{\rm trun}$.
The simulated blue tilts are found to be seen more clearly in more massive
galaxies, which reflects the fact that more massive galaxies
at $z=0$ are formed from a larger number of dwarfs with stellar nuclei
formed at $z>z_{\rm trun}$.
The half-number radii ($R_{\rm e}$) of GCSs, velocity dispersions
of GCSs ($\sigma$), and their host galaxy masses ($M_{\rm h}$)
are found to be correlated with one another 
such that $R_{\rm e} \propto {M_{\rm h}}^{0.57}$
and $\sigma  \propto {M_{\rm h}}^{0.32}$.
Based on these results, we discuss the link
between hierarchical merging histories of galaxies
and the physical properties of MPCs,
the origin of the $L-Z_{\rm gc}$ relation,
and non-homology of GCSs. 
\end{abstract}

\begin{keywords}
globular clusters: general --
galaxies: star clusters --
galaxies:evolution -- 
galaxies:stellar content
\end{keywords}

\section{Introduction}

A growing number of observational
studies of 
globular cluster systems (GCSs) 
have revealed interesting correlations 
between physical properties of GCSs and those of their
host galaxies (see Brodie \& Strader 2006 for a recent review).
For example,
%the radial density  profiles of GCSs are 
%observed to be shallower for lower luminosity galaxies
%(e.g., Ashman \& Zepf 1998).
Strader et al. (2004) found  that
mean colors of metal-poor globular clusters (GCs)
correlate  with luminosities of their host galaxies
and accordingly  suggested  that mean GC metallicities
($Z_{\rm gc}$) depend on luminosities ($L$)
of their host galaxies such that $Z_{\rm gc} \propto L^{0.15 \pm 0.03}$.
Similar results ($Z_{\rm gc} \propto L^{0.16 \pm 0.04}$) 
were found using the Advanced Camera for Surveys on the
{\it  Hubble Space Telescope (HST)} by Peng et al. (2006; P06). 
Brodie \& Strader (2006) have suggested that the relation was
steeper in the past. They argue that more enriched metal-poor GCs
form
first in low mass objects (from early
collapsing peaks at high redshift). These low-mass building
blocks merge, ultimately forming  
the massive galaxies of high density regions today. Thus the
slope, particularly at the high mass end, 
flattens over time.
%Forbes (2005) has recently shown that the specific frequency ($S_{\rm N}$)
%is higher for lower luminosity early-type galaxies based on
%{\it HST} ACS data.

Although such observed correlations between the physical  properties
of GCSs and their
host galaxies have been suggested to contain
fossil information of galaxy formation and evolution
(e.g., Harris 1991; Forbes \& Forte 2001; West et al. 2004; Brodie \& Strader 2006),
the details remain largely unclear
owing to the lack of theoretical and numerical studies
of GCSs.
Based on dissipationless numerical simulations of
major galaxy mergers with GCs,
Bekki \& Forbes (2006) first demonstrated that the GCS radial density
profiles dependent on galaxy luminosity
can be  understood in terms of 
the number of major merger events experienced. 
The dependence of GC destruction 
on host galaxy mass 
has also been suggested to be important for 
understanding the inner radial density profiles 
(Baumgardt 1998; Vesperini et al. 2003; Capuzzo-Dolcetta 2006).
%Correlations between kinematical and chemical properties of GCSs
%and those of their host galaxies have not been discussed
%in galaxy-scale numerical simulations.

The formation of GCs in low-mass dark matter halos
at high redshifts have been investigated by numerical and theoretical studies
(e.g., Broom \& Clarke 2002; Mashchenko \& Sills 2005).
Furthermore, the 
physical properties of GCSs have recently been discussed 
by several authors based on hierarchical galaxy formation scenarios where
GC formation occurs at high redshifts 
(e.g., Beasley et al. 2002; Kravtsov \& Gnedin 2005;
Santos 2003; Bekki 2005; Rhode et al. 2005; Yahagi \& Bekki 2005;
Moore et al. 2006; Bekki et al. 2006; Bekki \& Yahagi 2006).
For example,  Beasley et al. (2002) first demonstrated that 
the observed bimodal color distributions of GCSs in elliptical galaxies
can be reproduced by a semi-analytic galaxy formation
model. 
They also presented GC colour vs galaxy magnitude
relations, however they did not fully reproduce the observed
trend.
Although these previous works discussed some physical properties
of GCSs in galaxies, they did not explore in detail.  
The correlations between 
GCSs and those of their host galaxy. 
%(e.g., the $L-Z_{\rm gc}$ relation). 
Thus it remains unclear how the physical properties of GCSs
evolve in a hierarchical merging cosmology.

The properties of {\it metal-poor GCs (MPCs)} ([Fe/H] $\sim$
$-1.5$) such as their very
old ages, low metallicities and extended spatial distribution all
suggest formation at early times, when the low mass building blocks
of galaxies formed. 
The purpose of this current paper is to investigate
the physical properties of MPCs and their scaling relations
with their host galaxies based on
high-resolution cosmological simulations that
follow realistic merging and accretion histories of galaxies.
We investigate the structural, kinematical, 
and chemical  properties of MPCs.
We also present several predictions for 
the expected correlations between properties
of MPC systems (such as half-number radii,
effective surface number densities, and velocity dispersions). 
Since the present model is based on dissipationless
simulations of MPC formation within halos at
high redshifts, the origin of {\it metal-rich GCs (MRCs)} ([Fe/H]
$\sim$ $-0.5$) formed somewhat later during
dissipative merging will not be discussed.
We plan to investigate MRCs in a future paper
by combining the present high-resolution
cosmological simulation with semi-analytic models
similar to those used by Beasley et al. (2002).

 The plan of this paper is as follows: In the next section,
we describe our numerical models of MPC formation and their
assumed initial properties.
In \S 3, we 
present our numerical results
on (i) $L-Z_{\rm gc}$ relations, (ii) scaling relations
between properties of MPCs, and (iii) metallicity-magnitude
relations for luminous MPCs in massive galaxies (the so-called `blue-tilt'). 
In \S 4, we discuss our results.
We summarize our  conclusions in \S 5. Appendices explore scaling
relation dependency on M$_S$/L and z$_{trun}$. 
Throughout this paper,  GCSs 
in our simulations are composed only of MPCs.

\section{The Model}

\subsection{Identification of MPCs}

We simulate the large scale structure of GCs  
in a $\Lambda$CDM Universe with ${\Omega} =0.3$, 
$\Lambda=0.7$, $H_{0}=70$ km $\rm s^{-1}$ ${\rm Mpc}^{-1}$,
and ${\sigma}_{8}=0.9$ 
by using the Adaptive Mesh Refinement $N-$body code developed
by Yahagi (2005) and Yahagi et al. (2004), 
which is a vectorized and parallelized version
of the code described in Yahagi \& Yoshii (2001).
We use $512^3$ collisionless dark matter (DM) particles in a simulation
with the box size of $70h^{-1}$Mpc and the total mass 
of $4.08 \times 10^{16} {\rm M}_{\odot}$. 
We start simulations at $z=41$ and follow it until $z=0$
in order to investigate physical properties
of old GCs outside and inside virialized dark matter halos. 
We used COSMICS (Cosmological Initial Conditions and
Microwave Anisotropy Codes), which is a package
of fortran programs for generating Gaussian random initial
conditions for nonlinear structure formation simulations
(Bertschinger 1995, 2001).

Our method of investigating GC properties is described as follows.
Firstly, we select virialized dark matter subhalos at a
truncation redshift $z=z_{\rm trun}$ 
by using the friends-of-friends (FoF) algorithm (Davis et al. 1985)
with a fixed linking length of 0.2 times the mean DM particle separation.
%Details of the truncation redshift 
%The reasonable values of $z_{\rm trun}$ and the reasons for our
%introducing $z_{\rm trun}$ 
%are described later.
The minimum particle number $N_{\rm min}$ for halos is set to be 10.
For each individual virialized subhalo
with a half-mass radius of $R_{\rm h}$,
particles within $R_{\rm h}/3$ are labeled 
as ``GC'' particles and are considered to be old MPCs. 
This procedure for defining GC particles
is based on the assumption that energy dissipation via radiative cooling
allows baryons to fall into the deepest potential well of dark-matter halos
and finally to be converted into GCs.
The value of the truncation radius ($R_{\rm tr,gc}$ =  $R_{\rm h}/3$)
is chosen, because the size 
of the old GCs in
the Galactic GC system 
(i.e., the radius within which most Galactic old GCs are located)
is similar to $R_{\rm h}/3$ of the dark matter
halo in a dynamical model of the Galaxy (Bekki et al. 2005).
We assume that old MPC formation is truncated
completely after $z=z_{\rm trun}$ and investigate the dependences of
the results on $z_{\rm trun}$. Physical motivation for  the truncation
of GC formation  
is described later.

Secondly, we follow  GC particles
formed before $z=z_{\rm trun}$ until $z=0$ and thereby
derive locations $(x,y,z)$ 
and velocities $(v_{\rm x},v_{\rm y},v_{\rm z})$
of GCs at $z=0$.
We then identify virialized halos at $z=0$ with the FoF algorithm
and investigate whether each GC is within a halo.
If GCs are found to be within a halo, the mass of the host halo
($M_{\rm h}$)
and physical properties of the GCS  
are investigated.
If a GC is not in any halo, 
it is regarded as an intergalactic GC. 
The number fraction of these depends on $z_{\rm trun}$
but is typically less than 1 \% (e.g., $\sim$ 0.3 \% for $z_{\rm
trun}=10$). We don't consider intergalactic GCs further in this
current paper.

Thus, the present simulations enable us to investigate
the physical properties only for old MPCs 
due to the adopted assumption of collisionless simulations. 
The physical properties of MRCs which may form later 
during dissipative merger events (e.g., Ashman \& Zepf 1992) are
not investigated.

\begin{figure*}
\psfig{file=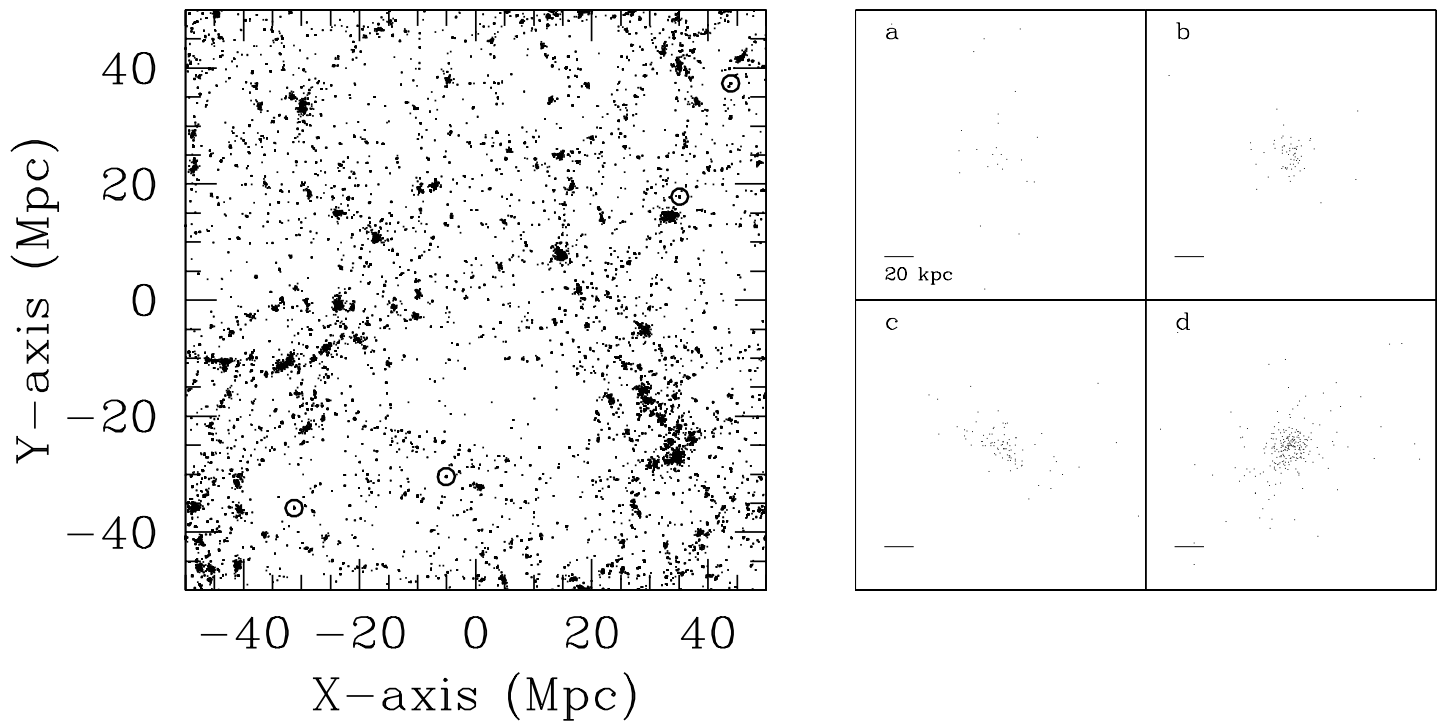,width=19.0cm}
\caption{ 
Spatial distributions of all MPCs projected onto the $x$-$y$ plane
in the large-scale cosmological simulation  (left)
and spatial distributions of MPCs within the selected four galaxy-scale
halos with similar masses (right) for the model with $z_{\rm trun}=10$.
The halo masses $M_{\rm h}$ are 
$9.3 \times 10^{12} {\rm M}_{\odot}$,
$9.6 \times 10^{12} {\rm M}_{\odot}$,
$6.1 \times 10^{12} {\rm M}_{\odot}$,
and $8.1 \times 10^{12} {\rm M}_{\odot}$,
for the upper left (a), the upper right (b),
the lower left (c), and lower right (d), respectively.
The four circles shown by solid lines represent
the locations of the four halos in the left panel. 
The horizontal bar is 20 kpc. 
The total number of MPCs range from 31 to 249
in the four halos, which reflects the different
merging histories with MPCs formed before $z_{\rm trun}$ =10.
} 
\label{Figure. 1}
\end{figure*}

\subsection{Initial properties of MPCs in galaxy-scale halos}

\subsubsection{Radial density profiles}

We assume that the initial radial profiles of GCSs ($\rho (r)$) in
subhalos at $z=z_{\rm trun}$ are the same as those described by 
the universal ``NFW'' profiles
(Navarro, Frenk \& White 1996) with 
$\rho (r) \propto r^{-3}$ in their outer parts.
The mean mass of subhalos at $z=z_{\rm trun}$ 
in the present simulations is roughly
similar to the total mass
of dwarf  galaxies today. 
Minniti et al. (1996) found that the projected ($R$) density
profiles of GCSs in dwarfs is approximated as $\rho (R) \propto R^{-2}$, which
translates roughly to  $\rho (r) \propto r^{-3}$
using a canonical conversion formula from
$\rho (R)$ into $\rho (r)$ (Binney \& Tremaine 1987).
Therefore, the above $r^{-3}$ dependency can be regarded as reasonable.
Below we mainly show the fiducial model with $\rho (r)$ similar
to the NFW profiles and $R_{\rm tr,gc}=R_{\rm h}/3$.

Although we base our GC models
on observational results of GCSs at z=0,
we  can not confirm whether 
the above 
$\rho (r)$ and  $R_{\rm tr,gc}$ values 
are reasonable 
for  GCSs in low-mass subhalos at z $\ge$ 6 owing to the  lack of
observational data for GCSs at high redshifts.
Although our present results on the $L-Z_{\rm gc}$ relation
and the blue tilt  at $z=0$ do not depend
strongly on the initial distributions of MPCs, 
structural properties of GCSs do depend on the spatial distribution.
For example, if we choose smaller  $R_{\rm tr,gc}$ at $z=z_{\rm trun}$,
the final projected number distributions of GCSs becomes more 
compact. 
%Thus the observed properties of GCSs 
%give some constraints on the birth places of MPCs within galaxies
%at high redshifts, if they are compared with the present simulation
%results. 

\subsubsection{The $L$--$Z_{\rm gc}$ relation}

In order to investigate the mean metallicities of MPCs 
in galaxy-scale halos at $z=0$, we need to allocate
initial metallicities to all GCs formed before $z=z_{\rm trun}$.
We assume that GCs in a halo at $z_{\rm trun}$ have 
identical metallicities of $Z_{\rm gc}$ and that 
the metallicities are a function of the total stellar mass (or luminosity)
of the halo. 
Lotz et al. (2004) have found that
$Z_{\rm gc} \propto L^{ {\alpha}_{\rm gc} }$, 
where ${\alpha}_{\rm gc} = 0.2$ for $B-$band luminosities in dEs
and dE,Ns.
This relation is  similar to $Z_{\rm gc} \propto L^{0.15}$ discovered 
by Strader et al. (2004).
Dekel \& Silk (1980) demonstrated  
that the stellar metallicities ($Z$) of dwarf galaxies 
embedded in massive dark matter halos
correlate with luminosities ($L$) of the dwarfs such
that $L \propto Z^{2.7}$.
This translates to 
$Z_{\rm gc} \propto L^{ {0.37} }$ if GCs follow field stars in dwarfs.
Prompted by these theoretical and observational studies, 
we adopt the following relation at high redshift 
between [Fe/H] (or log $Z_{\rm gc}$)
of MPCs  and $L$:
\begin{equation} 
{\rm [Fe/H]}  = {\alpha}_{\rm gc} \times \log L + {\beta}_{\rm gc},   
\end{equation}
where ${\alpha}_{\rm gc}$ and $ {\beta}_{\rm gc}$ are 
set to be 0.5 and -6.0 in the present study. We show the models
with these values, because they are more consistent with
the observed $L-Z_{\rm gc}$ relation.
% by P06 that clearly
%describes the values of ${\alpha}_{\rm gc}$ and  ${\beta}_{\rm gc}$.

In order to derive $L$ for halos with masses $M_{\rm h}$ at $z=0$,
we assume a mass-to-light-ratio ($M_{\rm h}/L$) 
dependent on $M_{\rm h}$. 
Recent observational studies based on galaxy luminosity
functions for luminous galaxies suggests that
$M_{\rm h}/L$ 
depends on $M_{\rm h}$ as 
$M_{\rm h}/L \propto {M_{\rm h}}^{0.33}$ (Marinoni \& Hudson 2002).
Dekel \& Silk (1986) proposed that $M_{\rm h}/L \propto L^{-0.37}$
for low-luminosity dwarf galaxies,
which can be interpreted as   $M_{\rm h}/L \propto {M_{\rm h}}^{-0.55}$.
These studies suggest that the $M_{\rm h}$-dependences of 
$M_{\rm h}/L$ are different between low- and high-luminosity
galaxies (see also Zaritsky et al. 2006). 
We thus adopt two different 
$M_{\rm h}$-dependent $M_{\rm h}/L$ ratios for galaxy-scale
halos in the present study.
For halos above a threshold halo mass of $M_{\rm h,th}$,
we adopt the following: 
\begin{equation} 
M_{\rm h}/L =  C_{\rm ML} \times { (\frac{M_{\rm h}}{M_{\rm h,th}}) }^{0.3},
\end{equation}
where $C_{\rm ML}$ is a constant and the value
of $M_{\rm h}/L$ at $M_{\rm h}=M_{\rm h,th}$.
For halos below a threshold halo mass of $M_{\rm h,th}$,
we adopt the following: 
\begin{equation} 
M_{\rm h}/L =  C_{\rm ML} \times { (\frac{M_{\rm h}}{M_{\rm h,th}}) }^{-0.5},
\end{equation}
We assume that  $C_{\rm ML}=10$ and $M_{\rm h,th}=10^{11} {\rm M}_{\odot}$
are reasonable values.
Since  the threshold mass $M_{\rm h,th}$ has not
yet been so precisely determined by observational studies 
(e.g., Zaritsky et al. 2006),
$M_{\rm h,th}$ can be a free parameter in our simulations.
We however confirm  that this  $M_{\rm h,th}$
for  a reasonable range  is not so important
as other parameters (e.g., $M_{\rm h}/L$ dependences). 
We thus mainly show the results of the models with
$M_{\rm h,th}=10^{11} {\rm M}_{\odot}$.

In calculating $Z_{\rm gc}$ from $L$ at $z=z_{\rm trun}$, 
we need to derive $L$ from $M_{\rm h}$ at $z=z_{\rm trun}$.
It is, however,  observationally unclear what 
a reasonable $M_{\rm h}$-dependent $M_{\rm h}/L$ is
for halos at $z=z_{\rm trun}$.
We accordingly investigate two extreme cases: 
One is that $M_{\rm h}/L= C_{\rm ML}$ for all halos at $z=z_{\rm trun}$
and the other is that the $M_{\rm h}$-dependence of  $M_{\rm h}/L$  
is the same between $z=0$ (equations 2 and 3) and $z=z_{\rm trun}$.
The total {\it stellar} mass ($M_{\rm s}$) 
of a halo is estimated from $M_{\rm h}$
by using the adopted  $M_{\rm h}$-dependent $M_{\rm h}/L$ 
and the following stellar-mass-to-light-ratio ($M_{\rm s}/L$).

It is highly likely that $M_{\rm h}$-dependences of
$M_{\rm h}/L$ are different between different redshifts,
because of (i) evolution (e.g., aging) of stellar populations
with different ages and metallicities and (ii)
evolution of baryonic mass fraction.
Although observational studies
on the redshift evolution of the ``Fundamental Plane'' 
of early type galaxies (e.g., van de Ven et al. 2003)
have  provided some clues to stellar population evolution 
it remains observationally unclear whether evolution 
of baryonic mass fraction  
is seen in galaxies: kinematical data sets for precisely  estimating
total masses of galaxies are currently unavailable for
very high-$z$ galaxies.
The present models, which do  not allow us to make explicit
predictions on  redshift evolution of $M_{\rm h}/L$,
have difficulties to determine
which of the above two is responsible for
the possible redshift evolution in
$M_{\rm h}$-dependences of $M_{\rm h}/L$. 
We therefore do not intend to discuss
the origin of the  redshift evolution in
$M_{\rm h}$-dependences of $M_{\rm h}/L$
in the present paper.

The stellar population synthesis models by Vazdekis et al. (1996)
for a Salpeter IMF show that $M_{\rm s}/L$ in the $V$-band
for stellar populations
with [Fe/H] = $-1.7$ (0.0) is 0.22 (0.39) for 0.5 Gyr and
2.03 (4.84) for 12.6 Gyr.  Although $M_{\rm s}/L$ depends strongly
on age and metallicity of the stellar population,
we use a constant $M_{\rm s}/L$ of 1 for all halos because
of the lack of information on stellar populations in
the simulated galaxy-scale halos.
%P06 also used an average luminosity-weighted $M_{\rm s}/L$
%for each galaxy in order to convert luminosities of galaxies into
%$M_{\rm s}$.  Since we compare the present simulations with
%observations by P06, it is reasonable to use a fixed $M_{\rm s}/L$
%in deriving $M_{\rm s}$. 
A discussion on the dependences on $M_{\rm s}/L$
is given in Appendix B.

Given the fact that
most of the virialized halos at $z_{\rm trun}$
are of low-mass ($< 10^{10} {\rm M}_{\odot}$) then the 
low-metallicities used to derive $M_{\rm s}/L$ should be reasonable.
We adopt $M_{\rm s}/L=1$, which is the average
value between 0.5 and 12.6 Gyrs in 
metal-poor stellar populations
{\it both} for  $z_{\rm trun}$ and $z=0$
for halos with different masses.
Although this adoption is more idealized,
it helps us to derive more clearly
the evolution of the $M_{\rm s}-Z_{\rm gc}$ relation
between  $z_{\rm trun}$ and $z=0$ without having complicated
dependences of $M_{\rm s}/L$ on ages and metallicities of stars
in halos (that can not be modeled directly in the present study). 
If we adopt a higher $M_{\rm s}/L$ (e.g., $\sim 4$),
the simulated $M_{\rm s}-Z_{\rm gc}$ relation at $z=0$ 
is shifted to lower metallicities for a given stellar mass but
with the same relation slope.

\subsubsection{GC Luminosity function}

The luminosity function  of GCs has been observationally suggested
to be universal and have the following form (e.g., Harris 1991):
\begin{equation} 
\Phi(M)= C_{0} \times \exp(-{(M-M^{0}_{V})}^2/2{\sigma_{\rm m}}^2),
\end{equation}
where  $C_{0}$ is a constant,   $M^{0}_{V}$ = $-7.23 \pm 0.23$,
and $\sigma_{\rm m}=1.25$ mag (Harris 1991).
We adopt this luminosity function and 
allocate luminosities to GC particles by generating
random numbers for $-11$ $\le$ $M_V$ $\le$ $-4$. We note
that our results are not very sensitive
to the choice of $M^{0}_{V}$ or $\sigma_{\rm m}$. 
%Although Richtler (2003) showed that  $M_{0}$ = $-7.48 \pm 0.11$ mag
%and $\sigma_{\rm m}=1.2$, we show only the results for the above LF
%of GCs  owing to the very weak dependences of the present
%results on  $M_{0}$ for a reasonable range of  $M_{0}$.

\subsection{Initial properties of galactic nuclei}

Previous studies have suggested that stellar galactic nuclei (GN)
in nucleated low-mass galaxies might be observed as very massive GCs
after the host galaxy of the GN is destroyed during merging with,
and accretion onto, more massive galaxies 
(e.g.,  Zinnecker et al. 1988; Freeman 1993; Bassino et al. 1994;
Bekki \& Freeman 2003; Bekki \& Chiba 2004).
The physical properties of very massive GCs in the Galaxy and
M31 (e.g., $\omega$ Cen and G1) are observed to be different
from those of ``normal'' GCs  (e.g., Freeman 1993).
%We accordingly distinguish  stripped SGN from stripped GCs in
%the present study and investigate separately physical properties
%of these SGN and GCs. 
We identify  stellar galactic nuclei particles (referred to
as GN particles) at $z=z_{\rm trun}$ and follow their evolution
until $z=0$.

The  ACS Virgo Cluster Survey by C\^ote et al. (2006)  recently found 
that the mass fraction of GN (stellar galactic nuclei)
to their host galaxies is typically
0.3\% in early-type galaxies.
We thus adopt the following relation between the total masses
of GN ($M_{\rm gn}$) and the total stellar masses of their hosts
$M_{\rm s}$:
\begin{equation} 
M_{\rm gn} = f_{\rm gn} \times M_{\rm s},
\end{equation}
where $f_{\rm gn}$ is  
considered to be a free parameter in the present study.
Recent observations of stellar populations of SGN in nearby
galaxies suggest that GN have a significant 
contribution from young stellar populations  
and thus suggest that GN are growing with time (e.g., Walcher et al. 2006). 
Therefore it is reasonable to adopt an $f_{\rm gn}$ value significantly
smaller than today's 0.3\%.
We generally adopt $f_{\rm gn}=0.001$ but 
also discuss the dependence of our present results on  $f_{\rm gn}$.

Lotz et al. (2004) and C\^ote et al. (2006) found
a correlation between $M_{\rm gn}$ (or GN luminosities)
and colors of  GN, which suggests a mass-metallicity relation 
for GN (i.e., more massive GN are more metal-rich).
This correlation and the above $M_{\rm gn}$-$M_{\rm s}$
relation implies that more luminous galaxies have
more metal-rich GN. 
Lotz et al. (2004) found that the metallicities
of field stars of dE,Ns ($Z_{\rm S}$) correlates
with galaxy luminosities ($L$) such that
$Z_{\rm S} \propto L^{0.4}$.
Kravtsov \& Gnedin (2005) showed that galaxies in their cosmological
simulations exhibit a strong correlation between the stellar mass
($M_{\rm S}$) and the average metallicity ($Z$)
of stars (described as $Z_{\rm S} \propto {M_{\rm S}}^{0.5}$).
Guided by these studies, we assume the following
relation between GN metallicities (log $Z_{\rm gn}$)
and luminosities of their hosts ($L$) :
\begin{equation} 
{\rm [Fe/H]}  = {\alpha}_{\rm gn} \times \log L + {\beta}_{\rm gn},   
\end{equation}
where 
${\alpha}_{\rm gn}$ and $ {\beta}_{\rm gn}$ are 
set to be the same as those adopted
for MPCs denoted as ${\alpha}_{\rm gc}$ and $ {\beta}_{\rm gc}$
(i.e., 0.5 and -6.0, respectively)  in the present study
for consistency. 
We discuss whether the adopted models  
can provide a physical explanations for the origin of
mass-metallicity relation of luminous GCs around
massive galaxies in \S 4.2.

\subsection{Truncation of MPC formation}

Semi-analytic galaxy formation models based on a
hierarchical clustering scenario
have shown that models with a
truncation of GC formation at $z \sim 5$
can better reproduce the observed color bimodality of GCs in
early-type galaxies (Beasley et al. 2002).
%However it remains unclear
%what physical mechanism is responsible for the truncation
%(or the severe suppression)  of GC formation in subgalactic halos
%virialized at later times.
%Using numerical simulations with 
%three dimensional radiation smoothed particle hydrodynamics,
%Susa \& Umemura (2004) demonstrated that star formation
%in less massive galaxies can be severely suppressed
%by ultraviolet background radiation in a reionized
%universe.
%This result may well suggest that the truncation epoch  of GC formation
%($z_{\rm trun}$) is closely associated with the epoch of reionization.
Recent cosmological simulations 
have demonstrated that truncation of GC formation 
by some physical mechanism (e.g., reionization) is necessary
to explain the  
very high specific frequency ($S_{\rm N}$)
in cluster ellipticals,  structural properties of the Galactic 
old stars and GCs, and 
the mass-dependent $S_{\rm N}$ trend (Santos 2003;
Bekki 2005; Bekki \& Chiba 2005; Rhode et al. 2005;
Moore et al. 2006; Bekki et al. 2006).
Observational studies however have not yet found
any strong evidence for the truncation of GC formation
in GCSs of nearby galaxies.

If $z_{\rm trun}$ is closely associated with the completion
of cosmic reionization, then $z_{\rm trun}$ may well range
from 6 (Fan et al. 2003) 
to 20 (Kogut et al. 2003). 
Although we investigate models with different  $z_{\rm trun}$ (
i.e. 6, 10, and 15), here
we  present the results of the model with $z_{\rm trun}=10$.
This is because the latest observations
by Microwave Anisotropy Probe ({\it WMAP}) suggested
that the epoch of reionization is $z={10.9}_{-2.3}^{+2.7}$ (Page et al. 2006),
and because numerical simulations suggest that models with higher
$z_{\rm trun}$ ($\ge 10$) are more consistent with
observations of MPCs (e.g., Bekki 2005; Moore et al. 2006). 

\begin{figure}
\psfig{file=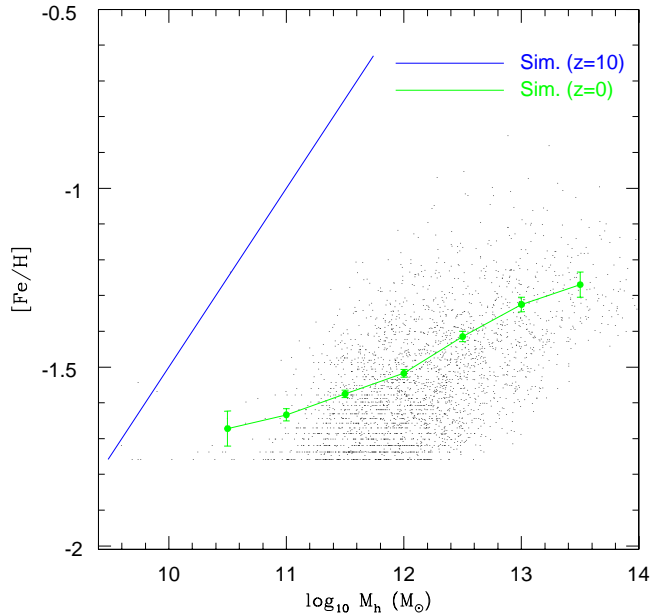,width=8.5cm}
\caption{ 
Mean metallicities ([Fe/H]) of metal-poor GCSs 
in simulated galaxies 
as a function of the total halo masses of the galaxies
(${\log}_{10} {\rm M}_{\rm h}$) for the model with $z_{\rm trun}=10$.
Each small dot represents a galaxy with a GCS  at $z=0$.
%Here we use a constant $M/L$ at $z=10$.
The thin blue line represents the initial 
relation (z = 10) between $Z_{\rm gc}$ and  $M_{\rm h}$,
where $Z_{\rm gc}$
is the mean metallicity of their MPCs.
The thin green line represents the mean value of 
the simulated relation today (z=0).
%[Fe/H]for different ${\log}_{10} {\rm M}_{\rm h}$.
}
\label{Figure. 2}
\end{figure}

\begin{figure}
\psfig{file=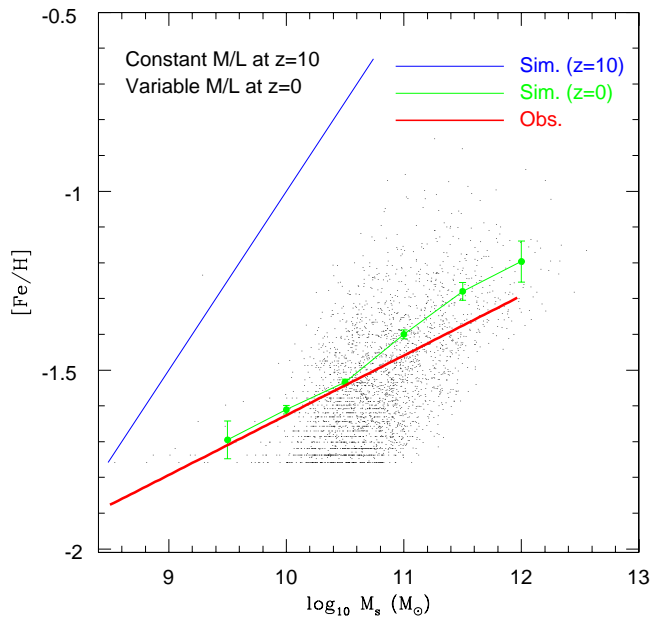,width=8.5cm}
\caption{ 
Mean metallicities ([Fe/H]) of metal-poor GCSs 
in simulated galaxies 
as a function of the total stellar masses of the galaxies
(${\log}_{10} {\rm M}_{\rm s}$) for the model with $z_{\rm trun}=10$.
Each small dot represents a galaxy with a GCS  at $z=0$.
Here we use a variable $M/L$ at  $z=0$ and a constant $M/L$ at $z=10$.
The thin blue line represents the initial $Z_{\rm gc} \propto L^{0.5}$ relation,
where $L$ is the total luminosity of galaxies and $Z_{\rm gc}$
is the mean metallicity of their MPCs.
The thin green line represents the mean value of 
the simulated relation today (z=0).
%[Fe/H]for different ${\log}_{10} {\rm M}_{\rm s}$.
For comparison, the observed $L-Z_{\rm gc}$ relation
of $Z_{\rm gc} \propto L^{0.16}$ (P06)  is shown by a thick red
line. 
}
\label{Figure. 3}
\end{figure}

\subsection{Galaxy-scale halos}

We simulate the structural, kinematical, and 
chemical properties of 
114596 GCs formed in 25096 low-mass halos  before $z_{\rm
trun}$ = 10. 
The physical properties of GCSs in 2830 halos 
with  masses $4.0 \times 10^9 {\rm M}_{\odot}$ $<$ $M_{\rm
h}$ $<$ $6.5 \times 10^{14} {\rm M}_{\odot}$ at $z=0$ are analyzed
and their correlations with host properties are investigated.
Since we are interested in the GCS properties in {\it galaxy-scale
halos} we only show the results for GCSs in halos with 
masses $10^{11} {\rm M}_{\odot} 
\le M_{\rm h} \le  10^{13} {\rm M}_{\odot}$
(and with $1 \le N_{\rm gc} \le 246$). 
%We thus do not analyze GCS properties in group-scale
%and cluster-scale halos ($10^{13} {\rm M}_{\odot}$ $<$ $M_{\rm h}$)
%in the present study.

We investigate the relations between three 
properties of GCSs
with $N_{\rm gc} \ge 4$, i.e. the half-number radii of projected
GC distribution ($R_{\rm e}$), the surface number densities
at $R_{\rm e}$ ($I_{\rm e}$), and the line-of-sight velocity 
dispersion for all MPCs in each halo ($\sigma$).
We derive these three properties, because they are observationally
feasible to measure. 
Since $M_{\rm s}/L$ is fixed at a constant value of 1 
in the present study,
the simulated $M_{\rm s}-Z_{\rm gc}$ relation is virtually
the same as the $L-Z_{\rm gc}$ relation that is often
discussed. The dependences of the present results on 
%(e.g., scaling relations and the $M_{\rm s}-Z_{\rm gc}$  relation)  on
$z_{\rm trun}$ is important, we show them in Appendices B and C.

\begin{figure}
\psfig{file=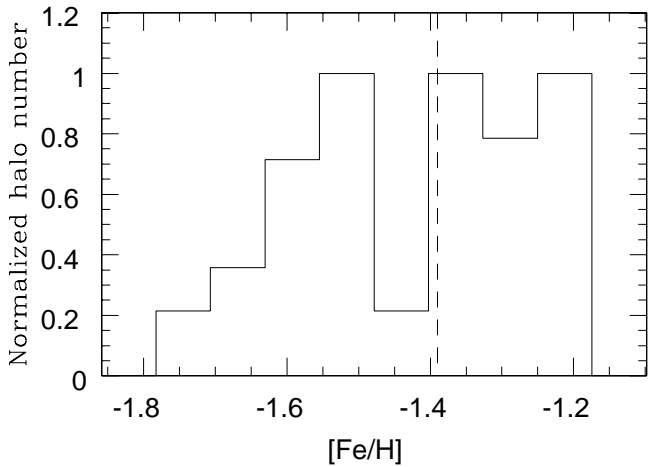,width=8.5cm}
\caption{ 
The metallicity distribution function (MDF) of GCSs from
progenitor low-mass halos
at $z=10$ that forms a galaxy-scale halo of 
$M_{\rm h}=5.0 \times 10^{12} {\rm M}_{\odot}$
at $z=0$.
This halo 
with a mean $Z_{\rm gc}=-1.39$ (dashed line) at $z=0$
originates from a low-mass halo
of $M_{\rm h}=4.0 \times 10^{10} {\rm M}_{\odot}$
at $z=10$ with $Z_{\rm gc}=-1.20$.
%The  dashed line represents the
%final (at $z=0$)
%$Z_{\rm gc}$ of the GCS  in the galaxy-scale halo.
The GCS 
at $z=0$ forms via hierarchical  merging and accretion, between
$z=10$ and $z=0$,   
of 74 low-mass halos each with a GCSs containing
a range of metallicities. 
%The dip at [Fe/H]$\sim -1.45$ is due to a small
%number of halos that have GCSs with  [Fe/H]$\sim -1.45$:
%Note that this is not a MDF  
%of all MPCs in this galaxy (but a MDF of GCSs of  its
%progenitor halos).
%Owing to the lower $Z_{\rm gc}$   of GCSs in merging/accreting
%low-mass halos,  the final $Z_{\rm gc}$  of the galaxy-scale
%halo becomes lower than the initial one.
}
\label{Figure. 4}
\end{figure}

\begin{figure}
\psfig{file=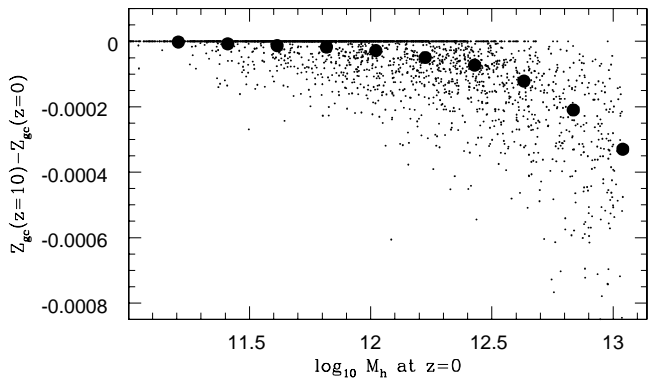,width=8.5cm}
\caption{ 
The metallicity difference
between GCSs in galaxy-scale halos at $z=0$ and
those in their progenitor halos at $z=10$
($\delta Z_{\rm gc} = Z_{\rm gc} (z=10)-Z_{\rm gc} (z=0)$)
with halo masses ($M_{\rm h}$) at $z=0$.
Small dots represent GCSs in galaxy-scale halos
and big filled circles  are the mean values of 
$\delta Z_{\rm gc}$.
% at each $M_{\rm h}$ bin.
}
\label{Figure. 5}
\end{figure}

\begin{figure}
\psfig{file=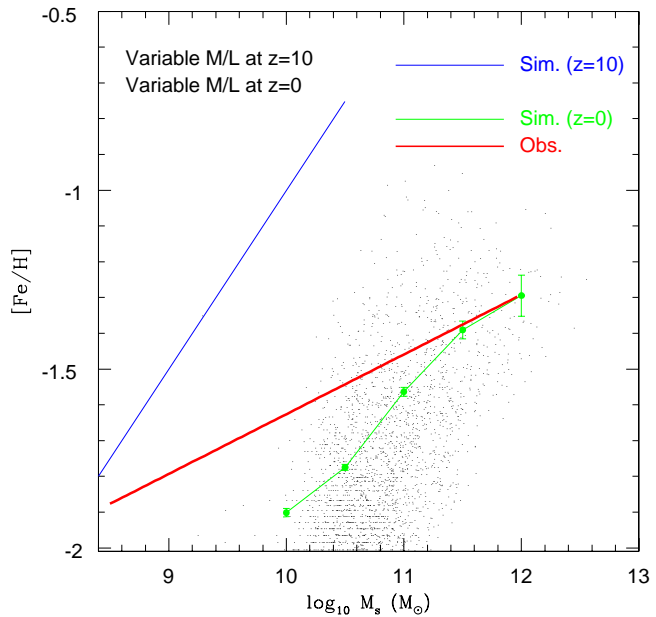,width=8.5cm}
\caption{ 
The same as Figure 3 but for the model with $z_{\rm trun}=10$ and
variable $M/L$ both at $z=0$ and at $z=10$.
}
\label{Figure. 6}
\end{figure}

\section{Result}

\subsection{The $L-Z_{\rm gc}$ relation}

Fig. 1 shows the large-scale ($\sim 50$ Mpc) distribution of
MPCs at $z=0$ and MPCs within
galaxy-scale halos with similar halo masses 
($M_{\rm h} \sim 
8 \times 10^{12} {\rm
 M}_{\odot}$)
in the model with $z_{\rm trun}=10$.
It it clear from Fig. 1 that 
the spatial distribution of MPCs in galaxy-scale halos
are quite diverse.
% depending
%on the merging histories of the halos. 
For example, the GCS distribution in the halo with
$M_{\rm h}= 6.1  \times 10^{12} {\rm M}_{\odot}$
appears to be more  flattened than that in
the halo with $M_{\rm h}= 9.6  \times 10^{12} {\rm M}_{\odot}$.
The mean spatial density of MPCs within the central 50 kpc  
of  the halo with $M_{\rm h}= 8.1  \times 10^{12} {\rm M}_{\odot}$ 
is a factor of $\sim 8$ higher  than
that of the halo with $M_{\rm h}= 9.3  \times 10^{12} {\rm M}_{\odot}$.
These diverse GCS structures are due to the differences of
merging histories of halos (e.g., the number of low-mass subhalos 
that were virialized before $z_{\rm trun}$ which merged to formed a halo).
The mean metallicities and metallicity distribution
functions of GCSs in galaxies are also diverse
depending on their merging histories.

Fig. 2 shows how the initial $M_{\rm h}-Z_{\rm gc}$ relation
evolves between $z=10$ and $z=0$ for the model with $z_{\rm trun}=10$.
Although the dispersion in [Fe/H] at $z=0$ is significantly large for
a given metallicity bin,  the $M_{\rm h}-Z_{\rm gc}$ relation
at $z=10$ becomes flatter at $z=0$.
This result clearly indicates that the mean metallicities of
MPCs in galaxies correlate with the total masses of
their host galaxies and thus with those of dark matter halos.
This also demonstrates that the simulated
$L-Z_{\rm gc}$ (discussed later) 
results from the $M_{\rm h}-Z_{\rm gc}$ relation
rather than the adopted assumptions on $M_{\rm h}/L$. 
The dependences of this correlation on $z_{\rm trun}$ are given
in Appendix A.

Fig. 3 shows that the initially steep $L-Z_{\rm gc}$ relation
at $z=10$ becomes significantly flatter at $z=0$
due to hierarchical merging
of low-mass galaxies formed at high $z$.
Fig. 3 also shows that the dispersion in [Fe/H] for a given
mass range of $M_{\rm s}$ is quite large.
%which suggests mean metallicities  and metallicity distribution
%functions of MPCs in galaxies are significantly different
%between galaxies with similar stellar masses.
Fig. 3  demonstrates that the observed $L-Z_{\rm gc}$
relation (i.e.,  $Z_{\rm gc} \propto L^{0.16}$)
at $z=0$
can be reasonably well reproduced by the model
with an initial $L-Z_{\rm gc}$ relation
of $Z_{\rm gc} \propto L^{0.5}$ at $z=10$.
We find that such a steep initial $L-Z_{\rm gc}$ relation
at $z_{\rm trun}$ 
is indispensable for reproducing the very flat yet significant
correlation between $L$ and $Z_{\rm gc}$ at $z=0$.
The flattening of the initial $L-Z_{\rm gc}$ relation
can be seen also in models with different $z_{\rm trun}$
(see Appendix C)

Figs. 4 and 5 show why the initial
steep $L-Z_{\rm gc}$ relation becomes flatter
during the hierarchical growth of galaxies.
Fig. 4 shows that the final ($z = 0$) mean metallicity 
of the MPCs in a galaxy with $M_{\rm h}=5.0 \times 10^{12} {\rm M}_{\odot}$
is lower ($Z_{\rm gc}=-1.39$) than the initial mean
metallicity 
($Z_{\rm gc}=-1.20$) of its progenitor low-mass halo with 
$M_{\rm h}=4.0 \times 10^{10} {\rm M}_{\odot}$,
which is the most massive halo  among those forming
this galaxy at $z=10$.
This galaxy grows via merging and accretion of 74 low-mass
halos, all with GCSs of different  metallicities between $z=10$ and $z=0$.
Since the MPCs in the accreted ``building blocks''
are metal-poor,
 the metallicity of
the GCS in the final galaxy is lower than that of
the GCS of its progenitor halo.
There is thus a decrease in the mean metallicities of GCSs 
during  galaxy evolution  between $z_{\rm trun}$ and $z=0$
which is driven by merging and accretion of low-mass halos
and their lower metallicity GCSs. 

Fig. 5 shows the differences in mean metallicities ($Z_{\rm gc}$) between
GCSs in galaxies at $z=0$ and those in their progenitor low-mass
halos at $z=10$ (i.e. $\delta Z_{\rm gc} = Z_{\rm gc} (z=10)-Z_{\rm gc} (z=0)$).
Although the dispersion in  $\delta Z_{\rm gc}$ 
is quite large in more massive galaxies,
the absolute values of 
$\delta Z_{\rm gc}$ are significantly larger in more massive
galaxies. 
This means that more massive galaxies at $z=0$ have experienced
a larger number of merging and accretion events 
of lower-mass halos with lower $Z_{\rm gc}$, 
so that the final $Z_{\rm gc}$ at $z=0$ is lower.  
%to a larger extend than their original values at $z=10$.
Low-mass galaxies at $z=0$, on the other hand,
have not experienced so many mergers and accretion events
that act to increase their masses and decrease their $Z_{\rm
gc}$.
Such galaxies are of low-mass 
because they have only grown slightly through merging/accretion 
since their virialization at $z=10$.
As a result of this and forming relatively many GCs in the
progenitor halo, 
their $Z_{\rm gc}$ does not change so much from
their original values at $z=10$.
Thus the  mass-dependence  of $\delta Z_{\rm gc}$
is the main reason for the flattening of the  $L-Z_{\rm gc}$
relation between $z_{\rm trun}$ and $z=0$.

Fig. 6 shows that if the $M/L$ ratios in galaxies are variable at
$z=10$ and $z=0$, then the simulated $Z_{\rm gc} \propto L^{0.3}$
relation is not a good fit to the observed one.  These results,
combined with those in Fig. 3, demonstrate that the model with a
constant $M/L$ at $z=10$ can better reproduce the observed flat
$L-Z_{\rm gc}$ relation for a given initial $L-Z_{\rm gc}$
relation.  Given the fact that both the $M_{\rm h}$-dependence of
$M/L$ and $L-Z_{\rm gc}$ relation at $z=10$ are observationally
unclear, there are two possible interpretations for the derived
steeper $L-Z_{\rm gc}$ relation at $z=0$ in the variable $M/L$
models.  One possibility is that the variable $M/L$ is more reasonable, and
thus the initial $L-Z_{\rm gc}$ relation at $z=10$ should be much
flatter than the adopted one of $Z_{\rm gc} \propto L^{0.5}$. The
other possibility is that a constant $M/L$ is more reasonable and the
observed $M_{\rm h}$-dependence of $M/L$ at $z=0$ is achieved
during galaxy evolution between $z=10$ and $z=0$.

An initial $L-Z_{\rm gc}$ relation at $z=10$ is required to  
have $Z_{\rm gc} \propto L^{0.25}$ 
to explain the observed relation in a variable $M/L$  model,
because merging flattens the slope by a factor of 1.7
in this model.  This $L-Z_{\rm gc}$ relation  at $z=10$
is flatter than the luminosity-metallicity relation  
of low-mass galaxies 
($Z \propto L^{0.37}$)
predicted by Dekel \& Silk (1980).  
Given the fact that there are no observational constraints 
on an initial $L-Z_{\rm gc}$ relation at $z=10$,
we can not conclude whether constant $M/L$ models
with initially steep $L-Z_{\rm gc}$ relations are better than
variable $M/L$ models with less steep $L-Z_{\rm gc}$ relations 
at this stage.

The present simulations thus show two extreme cases of
$L-Z_{\rm gc}$  evolution depending on 
the assumed $M/L$ at $z=z_{\rm trun}$.
%and thereby suggest the importance of adopting
%reasonable $M/L$ at  $z=z_{\rm trun}$.
The present simulations however confirm that 
the initial $L-Z_{\rm gc}$ relation
of GCSs at $z=z_{\rm trun}$
becomes significantly flatter by $z=0$,  irrespective of
different (yet reasonable) initial $M_{\rm h}$-dependences
of $M/L$ and different $z_{\rm trun}$.
%(e.g., $Z_{\rm gc} \propto L^{0.3}$ at $z=0$
%for the variable $M/L$ model). 
These results thus imply that the  $L-Z_{\rm gc}$
relation evolves with $z$ in such a way that
the $L-Z_{\rm gc}$ relation becomes flatter at lower $z$.
%This evolution of the $L-Z_{\rm gc}$ relation with redshifts
%has been already suggested by Brodie \& Strader (2006).

\begin{figure}
\psfig{file=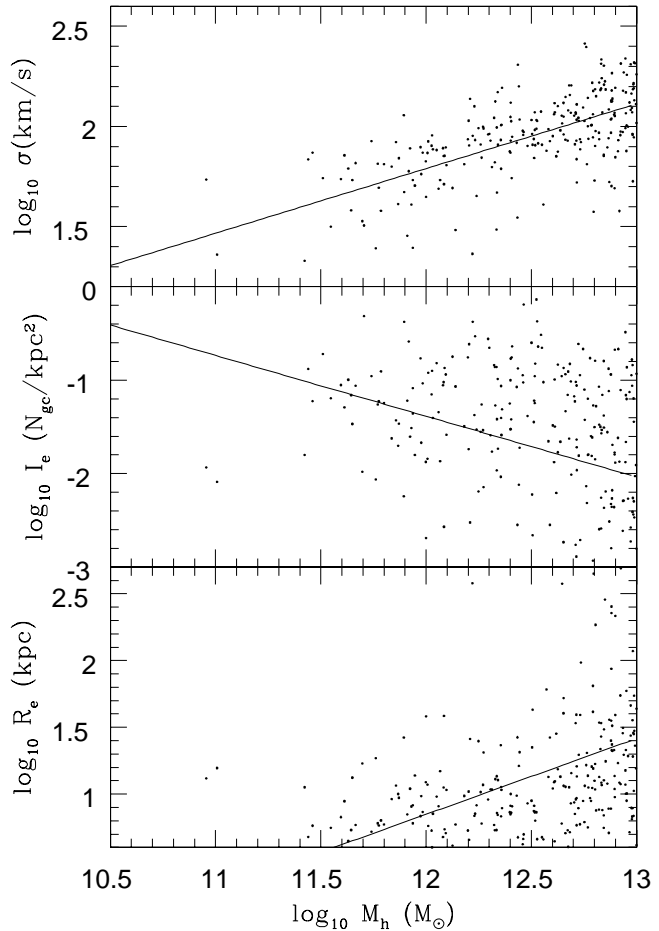,width=8.5cm}
\caption{ 
Velocity dispersions ($\sigma$, top),
effective surface number densities ($I_{\rm e}$, middle),
and half-number radii ($R_{\rm e}$, bottom) of MPCs  vs
the total halo masses of galaxies at $z=0$ for the model
with $z_{\rm trun}=10$. The solid line in each panel represents
the least-square fit to the simulation data.
}
\label{Figure. 7}
\end{figure}

\begin{figure}
\psfig{file=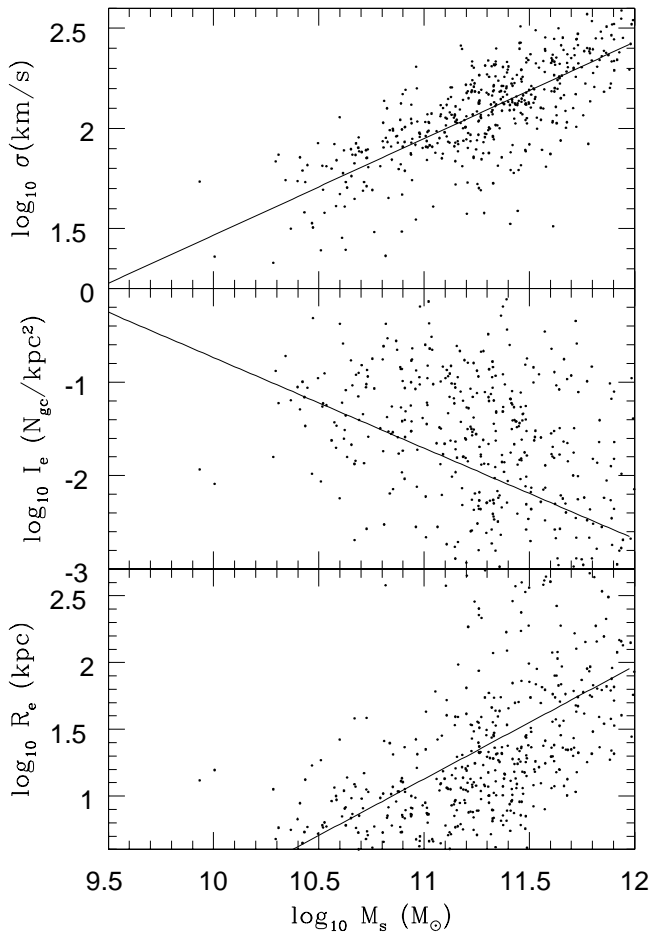,width=8.5cm}
\caption{ 
The same as Figure 7 but for the dependences on
total stellar masses of galaxies.
}
\label{Figure. 8}
\end{figure}

\begin{figure}
\psfig{file=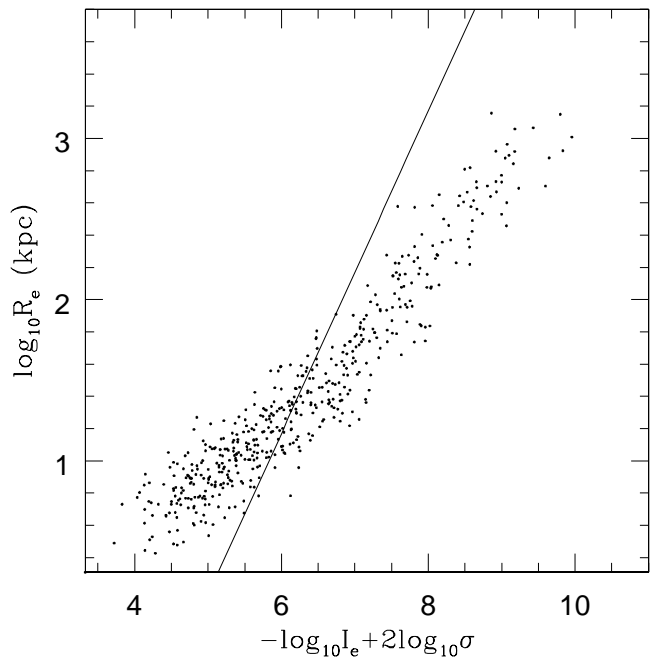,width=8.5cm}
\caption{ 
Half-number radii for {\it projected}
MPCs 
 as a function of the combination of $I_{\rm e}$ 
(surface-number density at  $R_{\rm e}$) and
$\sigma$ (line-of-sight velocity dispersion for MPCs)
for the model with $z_{\rm trun}=10$.
The solid line represents the ``virial plane''
defined as $R_{\rm e} = C_{\rm VP}  {I_{\rm e}}^{-1} {\sigma}^{2}$. 
Note that the distribution of GCSs deviates significantly
from the virial plane. Physical interpretations of this 
result is given in the main text.
}
\label{Figure. 9}
\end{figure}

\begin{figure}
\psfig{file=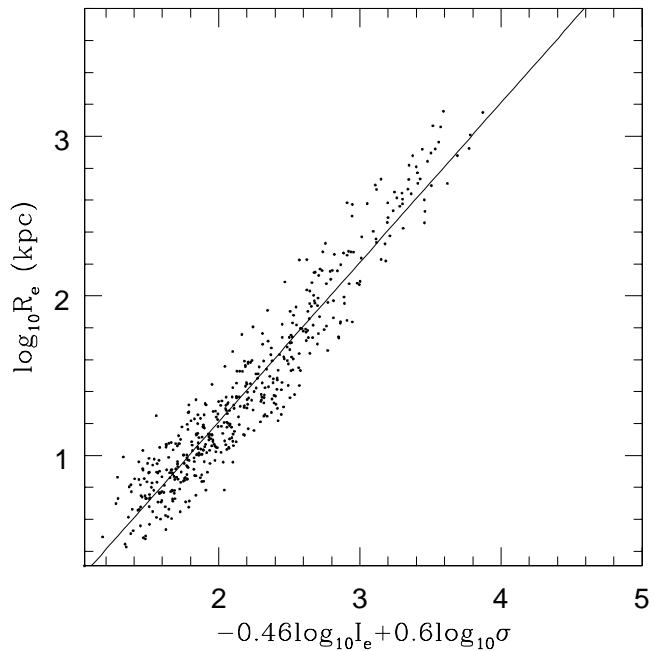,width=8.5cm}
\caption{ 
The same as Figure 9 but for the different abscissa.
The solid line represents the ``Fundamental Plane''
defined as $R_{\rm e} = C_{\rm FP}  {I_{\rm e}}^{-0.46} {\sigma}^{0.60}$.
The dispersion of the simulation data along this line
is  small, suggesting that GCSs are located
on a Fundamental Plane. Physical interpretations of this
result is given in the main text.
}
\label{Figure. 10}
\end{figure}

\subsection{Scaling relations}

Fig. 7  shows the scaling relations between
three projected properties of GCSs, 
$R_{\rm e}$ (half-number radii),
$I_{\rm e}$ (surface number densities at $R_{\rm e}$),
and $\sigma$ (line-of-sight velocity dispersion).
Fig. 7 demonstrates  that 
$\sigma$ and 
$R_{\rm e}$ strongly correlate with
$M_{\rm h}$ in such a way that  
$\sigma$ and
$R_{\rm e}$ are both larger for larger $M_{\rm h}$.
$I_{\rm e}$  however correlates weakly with 
$M_{\rm h}$ such that $I_{\rm e}$ is smaller for larger 
$M_{\rm h}$.
The least square fits to the simulation data show
that $R_{\rm e} \propto {M_{\rm h}}^{0.57}$,
$I_{\rm e}  \propto {M_{\rm h}}^{-0.65}$,
and $\sigma \propto {M_{\rm h}}^{0.32}$.
These scaling relations appear to be different from
those derived for dark matter halos (Kormendy \& Freeman 2004),
which suggests that GCSs in our models do not simply trace the
distributions of dark matter halos.
These scaling relations reflect the fact that MPCs in galaxy-scale
halos at $z=0$ originate from the central regions of low-mass
halos at $z=10$ and thus have different structures from those
of dark matter halos at $z=0$.

Fig. 8  shows how the properties 
of GCSs 
($R_{\rm e}$,  
$I_{\rm e}$  
$\sigma$)
correlate with total stellar masses ($M_{\rm s}$) in halos
and confirm  that these correlations are 
similar to those derived in Fig. 7.
The least square fits to the simulation data show
that $R_{\rm e} \propto {M_{\rm s}}^{0.85}$,
$I_{\rm e}  \propto {M_{\rm s}}^{-0.97}$,
and $\sigma  \propto {M_{\rm s}}^{0.48}$.
The derived $M_{\rm s}$-dependences are steeper than
the $M_{\rm h}$-dependences 
due to the adopted $M_{\rm h}$ variation with $M/L$ 
in galaxies at $z=0$.
These results suggest that 
future observational studies of GCSs
for a statistically significant number of GCSs 
can assess the validity of the present MPC formation
model.
%, if the observed  $M_{\rm s}$-dependences
%can be compared with the theoretically derived ones in
%the present study.
The dependency of the GCS scaling relations on $z_{\rm trun}$
are given and their physical meaning 
discussed in Appendix D.

Fig. 9 shows that the properties of GCSs 
(i.e., $R_{\rm e}$, $I_{\rm e}$, and $\sigma$)
in our models  do not
follow a relation for dynamical
systems in virial equilibrium 
(i.e.,  $R_{\rm e} = C_{\rm VP}  {I_{\rm e}}^{-1} {\sigma}^{2}$).
Given the fact that the simulated GCSs at $z=0$ are in dynamically
equilibrium,
this apparent deviation from the virial relation 
suggests structural and kinematical non-homology in GCSs, 
as observationally and theoretically
suggested for elliptical galaxies
(e.g.,  Djorgovski \& Davis 1987; Capelato et al. 1995). 
$R_{\rm e}$ and $\sigma$ of GCSs are significantly different from 
the half-mass radii and central velocity dispersions of dark matter
halos that host the GCSs. 
In  the present simulations,
MPCs within galaxy-scale halos at $z=0$
trace the particles  initially  within the central region of subhalos
virialized before $z=z_{\rm trun}$.
On the other hand,
dark matter halo particles at $z=0$ trace
all the particle from subhalos  virialized at every  $z$.
Therefore, $R_{\rm e}$ and $\sigma$ of GCSs
are quite different from the half-mass radii and central 
dispersions of dark matter halos that follow the virial relation.

Fig. 10 shows the best fit to
the three properties of 
$R_{\rm e}$, $I_{\rm e}$, and $\sigma$ for GCSs
(i.e.,  $R_{\rm e} = C_{\rm FP}  {I_{\rm e}}^{-0.46} {\sigma}^{0.60}$).
By assuming that 
$R_{\rm e} \propto  {I_{\rm e}}^{ {\alpha}_{\rm FP}  } 
\times {\sigma}^{  {\beta}_{\rm FP} }$,
the best values of ${\alpha}_{\rm FP}$ and ${\beta}_{\rm FP}$
are chosen such that the dispersion 
along  the assumed  line (or plane) with 
${\alpha}_{\rm FP}$ and ${\beta}_{\rm FP}$
become the smallest.
The derived relation of $R_{\rm e} = C_{\rm FP}  
{I_{\rm e}}^{-0.46} {\sigma}^{0.60}$ is quite different
from  the virial relation of 
$R_{\rm e} = C_{\rm VP}  {I_{\rm e}}^{-1} {\sigma}^{2}$,
which suggests structural and kinematical non-homology in
the simulated GCSs.
The more significant deviation of ${\beta}_{\rm FP}$
from the virial relation suggests that
the estimated line-of-sight velocity dispersions of GCSs 
can be quite different from (or significant smaller than) the central
velocity dispersions of dark matter halos required for
virial equilibrium. 
Since these three properties of
GCSs are feasible to derive observationally,
it is an interesting future study to compare these
simulated scaling relations with the corresponding
observational ones.

\begin{figure}
\psfig{file=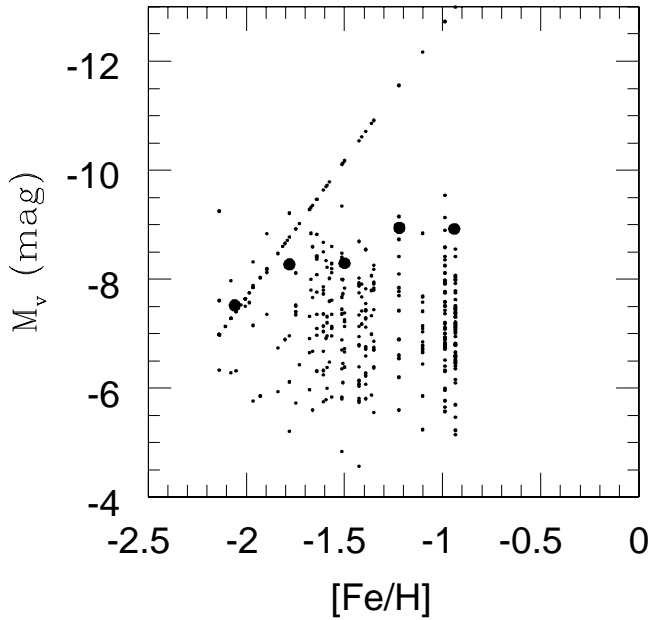,width=8.5cm}
\caption{ 
Distribution of MPCs in the $M_{\rm V}$-[Fe/H] plane for
the halo with $M_{\rm h}=3.0 \times 10^{13} {\rm M}_{\odot}$
and $N_{\rm gn}=62$
in the model with $f_{\rm gn}=0.001$.
% Here $N_{\rm gn}=62$ means that
Big dots represent the mean $M_{\rm V}$ for each of the five
metallicity bins.
This halo experienced merging/accretion of nucleated galaxies
62 times and thus now has 62 stripped nuclei in its GCS.
Note that a $M_{\rm V}$-[Fe/H] relation can be clearly
seen for MPCs with $-12$ $\le$ $M_{\rm V}$ $\le$ $-9$ (mag).  
This relation originates from the $L-Z_{\rm gn}$
relation of low-mass ``building blocks'' that formed this
halo.
The simulated $M_{\rm V}$-[Fe/H] relation of MPCs can
be compared with the observed ``blue tilt''.
The least-square fit to the five simulation data
shows $Z \propto L^{-2.01}$.}
\label{Figure. 11}
\end{figure}

\begin{figure}
\psfig{file=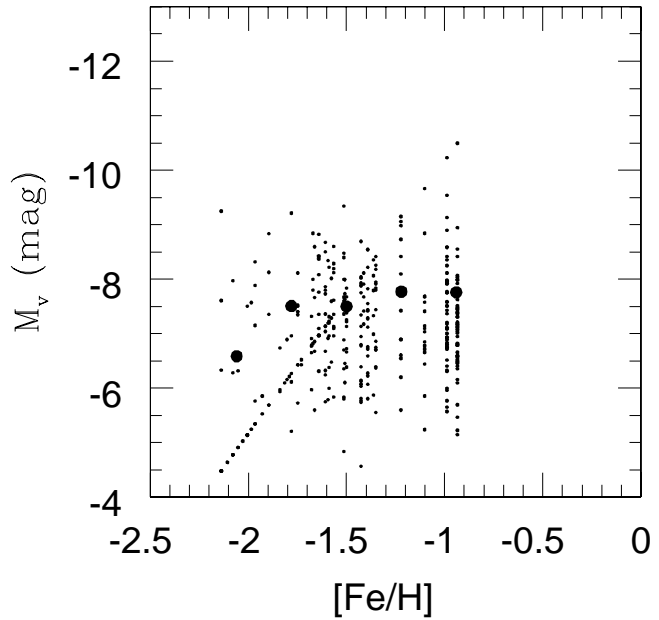,width=8.5cm}
\caption{ 
The same as Fig. 11 but for $f_{\rm gn}=0.0001$.
The least-square fit to the five simulation data
shows $Z \propto L^{-2.69}$.}
\label{Figure. 12}
\end{figure}

%\begin{figure}
%\psfig{file=f13.eps,width=8.5cm}
%\caption{ 
%The same as Figure 11 but for two small halos with similar masses. 
%A halo with $M_{\rm h}=2.9 \times 10^{12} {\rm M}_{\odot}$
%and $N_{\rm gn}=1$ (a) 
%and that with $M_{\rm h}=2.7 \times 10^{12} {\rm M}_{\odot}$
%and $N_{\rm gn}=14$ (b) are shown.
%}
%\label{Figure. 13}
%\end{figure}

\begin{figure}
\psfig{file=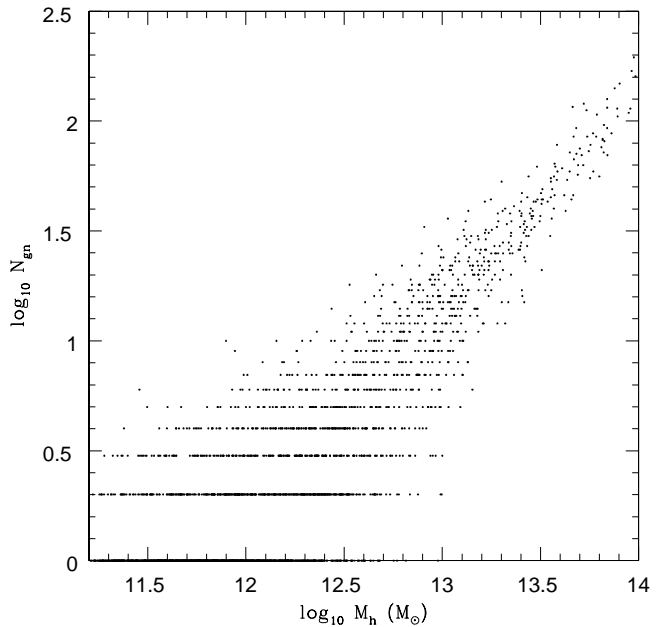,width=8.5cm}
\caption{
The number of galactic nuclei vs halo mass $M_{\rm h}$
for halos with $z_{\rm trun}=10$.
}
\label{Figure. 13}
\end{figure}

\subsection{Blue Tilts}

Next we describe the addition of accreted stellar galactic nuclei
(GN) to our simulated GCSs. 
Fig. 11 shows that the  GCS in a massive halo of $M_{\rm h}=3.0 \times
10^{13} {\rm M}_{\odot}$ 
% and no [Fe/H]-$M_{\rm V}$ relation 
%for the less luminous MPCs ($M_{\rm V} \ge -10$ mag)
for the model in which the fraction of the stellar galaxy mass in
nuclei is $f_{\rm gn}=0.001$.
The figure shows a clear [Fe/H]-$M_{\rm V}$ relation
in the more luminous MPCs ($M_{\rm V}<-9$ mag).
% similar to
%the observed ``blue tilt'' (e.g., Harris et al. 2006; Strader et al. 2006;
%Spitler et al. 2006).
The essential reason for the simulated blue tilt is
that this halo was formed from a large number of subhalos ($N_{\rm h} =62$)
that were virialized before $z_{\rm trun}$ and 
thus had GN that followed a mass-metallicity relation.
Whether or not the blue tilt in a halo at $z=0$  can be clearly seen
is determined by how many subhalos with GN particles 
merged to form the halo in the present simulations.
%It should be noted here that the observed blue tilts
%are not as clearly visible as seen in the present simulations. 
A least square fits to the simulation data show
that $M_{\rm V}  = -10.25 -1.24 \times {\rm [Fe/H]}$ 
(i.e., $Z \propto L^{2.01}$), which means that
the simulated blue tilt is significantly steeper than
the observed one (e.g., $Z \propto L^{0.55}$; Harris et al. 2006).

Fig. 12  shows that if $f_{\rm gn}=0.0001$, which is
an order of magnitude
smaller than the model in Fig. 11,
the blue tilt is much less clearly seen,
because low-luminosity GN particles can not be
distinguished from GC particles in the GCS.
The results 
in Figs. 11 and 12 imply that for a blue tilt to be clearly
seen at $z=0$, GN at $z=10$ should be as massive as $\sim  0.1$ \% of
their stellar components in their low-mass hosts.
We confirm that if $f_{\rm gn}=0.005$, a blue tilt
can be clearly seen but there are too many
very luminous ($M_{\rm V}<-12$ mag) MPCs in galaxy-scale halos.
The least square fits to the simulation data show
that $M_{\rm V}  = -8.22 -0.93 \times {\rm [Fe/H]}$ 
(i.e., $Z \propto L^{2.69}$), which is steeper than
the blue tilt in the model with $f_{\rm n}=0.001$
(and thus less consistent with observations).
These results suggest that the presence of blue tilts
gives some constraints on 
%the formation processes
%of GN at very high redshifts, if the origin 
%is closely associated with 
the destruction of low-mass nucleated
galaxies around galaxies between $z_{\rm trun}$ and $z=0$.
The dependences of the slopes of the simulated blue tilts on
the initial GC luminosity function are given in Appendix E.

%Although this result implies that the present models miss 
%some important ingredients of GC formation at high $z$,
%we do not have an idea about
%reasonable physical mechanisms by which
%the dispersions in [Fe/H] of GCSs in
%galaxies  can become much smaller. 
%We therefore discuss the origin of the observed smaller
%dispersion in [Fe/H] of GCSs in our future papers.

%Fig. 13 shows two halos with similar halo masses yet different merging
%histories (and thus different $N_{\rm gn}$). 
%Although it is almost impossible to identify the blue tilt for the halo 
%with $N_{\rm gn}=1$,  the blue tilt is discernible for 
%the halo $N_{\rm gn}=14$. 
%The halo with $N_{\rm gn}=1$ was formed both  from
%many subhalos formed after $z_{\rm trun}$ 
%(thus without MPCs) and from
%just one subhalo formed before  $z_{\rm trun}$
%(thus it has one GN and MPCs)  so that
%it does not show the blue tilt. 
%These results imply that
%galaxies with higher $S_{\rm N}$ of MPCs are more likely
%to show the blue tilt clearly.
%These results also suggest that not all galaxies will have
%the blue tilts in their GCSs.
Fig. 13 shows that more massive halos will have a larger 
number of GN particles within their GCSs and thus 
massive galaxies at $z=0$ are more likely to
have a contribution from disrupted nuclei to their observed blue
tilts.
The significantly smaller dispersion in $N_{\rm gn}$ 
for more massive halos with $M_{\rm h} > 10^{13} {\rm M}_{\odot}$
in Fig. 13 suggests that this effect is most pronounced for 
GCSs in the most luminous galaxies, i.e. those at the 
centers of groups and clusters.

The overall distribution of MPCs in the $M_{\rm V}$-[Fe/H]
plane does not reproduce well the observations
for GCSs with blue tilts (Harris et al. 2006; Strader et
al. 2006; Spitler et al. 2006).
%owing to the apparently large dispersion in [Fe/H] for MPCs with
%$M_{\rm V}>-9$ mag. 
The stripped nuclei, in this prescription,
are therefore unlikely to be the sole origin of the observed blue tilts
(although they may still make a small contribution).

\section{Discussion}
\subsection{The origin of the $L-Z_{\rm gc}$ relation}

The present numerical simulations are the first to demonstrate that
an initial steep $L-Z_{\rm gc}$ relation at high $z$
can become  significantly flatter due to hierarchical
merging of galaxies with GCSs. The slope of the
final $L-Z_{\rm gc}$ relation
GCSs at $z=0$ is similar to the observed one (Strader et al. 2004; P06)
in models with $z_{\rm trun}=10$ and constant $M/L$ at $z_{\rm trun}$.
C\^ote et al. (1998) 
investigated correlations between 
metal-poor GCSs and their host galaxy luminosities ($L$)
in their models of GCS formation via merging/accretion of 
MPCs from dwarfs with a power-law galaxy luminosity function (LF)
of slope $\alpha$.  
They showed that
there is a weaker correlation between 
the MPCs and $L$ in the models 
with a steep LF ($\alpha \sim -1.8$ in their Figure 7). 
Recent high-resolution cosmological simulations on the mass
function of low-mass halos have demonstrated that
the slope of the mass function is significantly steeper
at higher redshifts (e.g., Yahagi et al. 2004).
We suggest that the steeper mass function 
of halos at high redshifts
can also be an important factor
for the origin of the observed  
slope of  
the $L-Z_{\rm gc}$ relation.

The present study assumed that the initial $L-Z_{\rm gc}$ relation
at $z=z_{\rm trun}$ is similar to  the  $L-Z_{\rm S}$ relation
observed in Local Group dSphs  today (e.g., Dekel \& Silk 1984;
Lotz et al. 2004).
%  and to the $L-Z_{\rm gc}$ relation for the
%entire GC populations (including both MPCs and MRCs) in GCSs
%for early-type galaxies in the Virgo cluster (P06).
Such a steep $L-Z_{\rm gc}$ relation at $z=z_{\rm trun}$
is important
for the present model, because the initial  $L-Z_{\rm gc}$ relation
becomes significantly flatter between $z=z_{\rm trun}$ and $z=0$.
It is, however, observationally and theoretically unclear
how low-mass galaxies at high $z$  achieve 
such a steep relation and what physical mechanisms are
responsible. 
Kravtsov \& Gnedin (2005) found that galaxies in their cosmological
simulations exhibit a strong correlation between the stellar mass
($M_{S}$) and the average metallicity ($Z_S$)
of stars (described as $Z_S \propto {M_{S}}^{0.5}$)
similar to the observed one in dwarf galaxies (Dekel \& Woo 2003).
Their results imply that GCSs in low-mass galaxies at high $z$
could also have a steep $Z_{\rm gc}-M_{S}$ relation similar to
that of the field stars. 

It should be stressed here that the models with an initially steep
$L-Z_{\rm gc}$ relation at $z=z_{\rm trun}$
better reproduce observations only for  constant
$M/L$ at $z=z_{\rm trun}$: models with an initially flatter
$L-Z_{\rm gc}$ relation can also better explain observations
for variable $M/L$ at $z=z_{\rm trun}$,
though the required $L-Z_{\rm gc}$  relation was  not predicted
by previous theoretical studies. 
Owing to lack of observational data sets for 
the $L-Z_{\rm gc}$ relation at $z=z_{\rm trun}$,
it is currently difficult to make a robust conclusion
as to whether the constant $M/L$ models with
a steep $L-Z_{\rm gc}$ relation are better.

\subsection{The origin of the blue tilt}

Using the ACS on-board HST, Strader et al. (2006) 
and Harris et al. (2006) 
found that luminous blue GCs (i.e. MPCs) reveal a trend
of having redder colors in giant ellipticals. 
This ``blue tilt'' was interpreted as MPCs having a metallicity-luminosity
relation of $Z \propto L^{0.55}$. A possible exception was  
NGC 4472. Spitler et al. (2006) 
subsequently showed that
this trend is also true in the Sombrero spiral galaxy 
and may extend to lower GC
luminosities with a somewhat 
shallower slope than derived 
by Harris et al. (2006) and Strader et al. (2006). 
In each of these ACS studies the MRCs showed no corresponding trend. 
These observations require that any theoretical models
of GC formation 
should explain the origin of the 
metallicity-luminosity relation for individual
GCs, and also the apparent absence of a similar relation for MRCs
(e.g., Mieske et al. 2006).

Here we have demonstrated that
luminous MPCs in more massive  galaxies show metallicity-luminosity
relations in that brighter GCs have redder colors $-$ a trend  
qualitatively similar to the observed  blue tilts.
In our simulations the reason for the ``simulated blue tilts'' is that
luminous MPCs originate from stellar galactic nuclei (GN)
of the more massive  nucleated galaxies with
a luminosity-metallicity relation. 
The present study thus suggests that galaxies  which
experienced a larger number of accretion/merging events of
nucleated low-mass galaxies are more likely to show
a blue tilt. 
Since the frequency of accretion/merging events is 
generally higher in more massive galaxies, 
the blue tilts are more likely
to be present in high luminosity than low luminosity galaxies. 
Thus not all galaxies are expected to have blue tilts.
Since these simulated GCSs are composed both of stripped normal GCs
(with no initial mass-metallicity relation)
and of stripped GN with a  mass-metallicity relation, their
detection in observational data may depend on their relative numbers.
%The slopes of the blue tilts in GCSs are steeper
%(i.e.,  closer to the original slope of GN)
%for galaxies with larger $N_{\rm gn}$ (i.e., number
%of nucleated galaxies that are building blocks of
%galaxies at $z=0$) owing to the smaller dispersion caused
%by normal GCs in mass-metallicity relations (or color-magnitude
%diagrams).  

%Spitler et al. (2006) suggested that 
%if MRCs in a GCS  have color-magnitude relations with the slope
%similar to the observed blue tilt of the  MPCs,
%such a correlation with a smaller slope  can disappear in the noise
%for MRCs that have much higher metallicities.
%If the apparent absence of color-magnitude relations 
%is real,  it has implications for 
%the differences in formation processes between MPCs and MRCs
%(Harris et al. 2006; Strader et al. 2006; Spitler et al. 2006).
Although we find clear blue tilts in our simulations, it 
is not clear what role stripped galactic nuclei play in the observed blue 
tilts to date. For example, Spitler et al. (2006) find the blue tilt is 
not simply a feature of luminous MPCs but seems to extend down to 
luminosity function as far as is measurable. Furthermore the MPC 
luminosity function is well fit by a standard Gaussian, or t$_5$ 
function, even at the luminous end. Thus the tilt does not appear to be due 
to an additional population but rather a `tilting' of the existing 
MPC system. More deep data on a variety of systems are required to confirm 
these trends. A direct comparison between simulations and observational 
data needs to account for many factors, such as the number of GN vs bona 
fide GCs, contamination rates, the presence of dust and measurement errors 
(which tend to cause spreads in both magnitude and colour).

The observed colors and metallicities of
GN in dwarfs (Lotz et al. 2004; Mieske et al 2006) appear to suggest
most GN can not be metal-rich clusters as they have lower metallicities.
The apparent absence of a MRC color-magnitude
relation may therefore reflect the fact that
most MRCs do not originate from GN. Our simulated 
blue tilt, due to the 
stripped nuclei of (defunct) low-mass galaxies,  
is one of many possible scenarios for the origin of the blue tilt.
We plan to discuss the origin of the blue tilt
in a wider context of galaxy and GC formation in a
future paper.

\subsection{The Fundamental Plane of metal-poor GCSs}

The scaling relations between properties of elliptical
galaxies, such as ``the Fundamental Plane'' (FP), have long been
suggested provide valuable information on formation and evolution
histories of elliptical galaxies (e.g., Djorgovski \& Davis 1987).
Theoretical studies based mostly on
numerical simulations suggested that structural and kinematic
non-homology inferred from the FP are closely associated
with dynamics of galaxy merging, star formation histories 
dependent on galaxy masses, and collapse dynamics in elliptical
galaxy formation
(Capelato et al. 1995; Bekki 1998; Dantas et al. 2003).  
The present study has shown that the scaling relation
($R_{\rm e} = C_{\rm FP}  {I_{\rm e}}^{-0.46} {\sigma}^{0.60}$)  between
$R_{\rm e}$ (projected half-number radii), $I_{\rm e}$ (surface
number density at $R_{\rm e}$), and $\sigma$ (velocity
dispersion) of MPCs in our models 
is significantly different from that expected from the virial
theorem ($R_{\rm e} = C_{\rm VP}  {I_{\rm e}}^{-1} {\sigma}^{2}$). 
Although this is due partly to the way we estimate
$\sigma$ for a GCS in our models,
the significant difference between the exponents of the $I_{\rm e}$ terms
between the virial and the simulated relations
strongly suggests that the deviation has a physical meaning.

MPCs are assumed to form in the central regions
of low-mass halos that are virialized before $z=z_{\rm trun}$. 
The formation of MPCs is biased toward 
high density peaks of the  primordial matter
distribution in the adopted CDM model. 
As a natural result of this,
MPCs in galaxies at $z=0$ show structural and kinematical 
properties different from those of the background dark matter
halos, the scaling relation of which should be similar to the
virial relation. 
Recent numerical simulations with truncation of GC formation
have also demonstrated that the dynamical properties of MPCs can
be significantly different from those of dark matter halos
(Santos 2003; Bekki 2005; Moore et al. 2006).
Although physical explanations for the origin of the
slope of the FP in MPCs are yet to be provided,
we suggest that the presence of the FP for MPCs can be
one characteristic of biased formation of MPCs at high $z$.

A growing number of spectroscopic observations 
have revealed kinematic properties of GCSs,
such as radial profiles of velocity dispersions and
rotational velocities,
maximum $V/\sigma$,
and kinematical differences between MPCs and MRCs,
in early-type
galaxies (e.g.,  Kissler-Patig \& Gebhardt 1998; 
Cohen 2000; Zepf et al. 2001; C\^ote et al. 2001, 2003;
Beasley et al. 2004; Peng et al. 2004; Richtler et al. 2004; 
Pierce et al. 2006; Romanowsky 2006).
Recent photometric studies of GCSs by the  {\it HST} and ground-based
telescopes have also revealed structural properties of GCSs in
galaxies with different Hubble types
(e.g., Rhode \& Zepf 2004; P06; Spitler et al. 2006).
Accordingly it is worthwhile for future observational studies 
to investigate the FP of 
MPC systems.
%  and (2) the consistency of the observationally
%inferred slope of the FP with the predicted one. 
It is also an interesting observational study to investigate
correlations between $R_{\rm e}$ (or $I_{\rm e}$) of  MPCs 
and the total
luminosity of their host galaxy, because
such correlations are more  feasible to derive observationally
and thus can be compared more readily  with the predicted ones.

\section{Conclusions}

We have investigated the physical properties of metal-poor GCSs  in  galaxies 
using
high-resolution cosmological
simulations based on a $\Lambda$CDM model.  
We particularly investigated the $L-Z_{\rm gc}$
relation, the $M_{\rm V}-{\rm [Fe/H]}$ relation, 
and MPC scaling relations in galaxies with different masses. 
We assumed that 
(i) MPC formation is truncated at $z=z_{\rm trun}$,
(ii) galaxies at $z=z_{\rm trun}$ contain
MPCs and galactic nuclei (GN) 
with masses proportional to their host galaxies,
(iii) galaxies initially have a steep relation 
%of  $Z_{\rm gc} \propto L^{0.5}$ 
at $z=z_{\rm trun}$,
(iv) galaxies have a certain $M/L$ at $z=z_{\rm trun}$, 
and (v) MPCs have a luminosity function the same
as that observed today. 
We also assumed that both MPCs and GNs 
can be identified as  
MPCs in galaxies at $z=0$, if they are within virial radii
of the dark matter halos. 
We summarize our principle results of the models as follows.

(1) The original  $L-Z_{\rm gc}$ at $z_{\rm trun}$ becomes
significantly flatter by $z=0$ due to 
hierarchical merging of lower-mass galaxies. 
The original  $Z_{\rm gc} \propto L^{0.5}$ at $z=10$ 
in the model with a constant $M/L$ at $z_{\rm trun}$ 
becomes $Z_{\rm gc} \propto L^{0.15}$ at $z=0$,
which is consistent with the latest observational results. 
The origin of the flattening of the $L-Z_{\rm gc}$  relation
reflects the fact that
$Z_{\rm gc}$ of GCSs in more massive galaxies at $z=0$ is lower 
than the progenitor halos at  $z_{\rm trun}$
due to a large number of mergers and accretion
of low-mass halos of GCSs with lower  $Z_{\rm gc}$.
A flatter $L-Z_{\rm gc}$ relation at $z=10$
($Z_{\rm gc} \propto L^{0.25}$)  is required
to explain the observed relation
%($Z_{\rm gc} \propto L^{0.15}$) 
at $z=0$ in 
the models with a variable $M/L$.

(2) The final $L-Z_{\rm gc}$ relation at $z=0$ depends on
$z_{\rm trun}$ such that it is steeper for lower $z_{\rm trun}$.
%for a given original $L-Z_{\rm gc}$ relation  at $z=z_{\rm trun}$. 
Models with constant $M/L$ ratios (i.e., $M_{\rm h}/L=10$)
at $z=z_{\rm trun}$
can better explain the observed $L-Z_{\rm gc}$ relation  at $z=0$
for a given $z=z_{\rm trun}$ and the adopted  initial $L-Z_{\rm gc}$ relation 
of $Z_{\rm gc} \propto L^{0.5}$ 
at $z=z_{\rm trun}$,
though most models show a flattening of the original $L-Z_{\rm gc}$ 
relation.
The essential reason for the positive correlation 
between  $L$ and $ Z_{\rm gc}$ 
at $z=0$
is that the more massive galaxies at $z=0$ are formed from
hierarchical merging of a larger number of more massive
building blocks containing
GCs with higher metallicities. Thus merging does not 
completely erase the original $L-Z_{\rm gc}$ relation.

(3) MPC systems in these models 
clearly show  scaling relations 
between galaxy mass $M_{\rm s}$ (or halo mass $M_{\rm h}$),
$R_{\rm e}$, $I_{\rm e}$, and $\sigma$.
We find $R_{\rm e} \propto {M_{\rm s}}^{0.85}$,
$I_{\rm e,gc} \propto {M_{\rm s}}^{-0.97}$,
$\sigma \propto {M_{\rm s}}^{0.48}$ in the models
with $z_{\rm trun}=10$. 
Thus we find a ``Fundamental Plane'' for
MPCs, which can be described as 
$R_{\rm e} \propto  {I_{\rm e}}^{-0.5} {{\sigma}}^{0.6}$ 
in the models with $z_{\rm trun}=10$.
We note that such scalings are not consistent with the virial relation of 
$R_{\rm e} \propto  {I_{\rm e}}^{-1} {{\sigma}}^{2}$. The 
$R_{\rm e} - M_{\rm s}$
relation  depends on $z_{\rm trun}$, 
%such  that it is steeper
%for higher $z_{\rm trun}$,
which suggests that the scaling relation can give some 
constraints on $z_{\rm trun}$. We also find that
$R_{\rm e} \propto {M_{\rm h}}^{0.57}$
and ${\sigma}  \propto {M_{\rm h}}^{0.32}$.

(4) Luminous MPCs show a  correlation
between $M_{\rm V}$ and [Fe/H] if these
MPCs originate from nuclei of low-mass galaxies at high $z$.
The correlation, which is referred to as a ``blue tilt'',
can be more clearly seen in more massive galaxies
with $M_{\rm h} \sim 10^{13} {\rm M}_{\odot}$. This
is mainly because more massive galaxies are formed
from a larger number of nucleated galaxies virialized
by $z=z_{\rm trun}$. Observations of blue tilts only loosely
resemble our simulated ones suggesting stripped nuclei make a
small contribution at most.

\section*{Acknowledgments}
We are  grateful to the anonymous referee for valuable comments,
which contribute to improve the present paper.
We thank J. Strader and L. Spitler for useful discussions.
K.B. and  D.A.F. acknowledge the financial support of the Australian Research 
Council throughout the course of this work.
H.Y. acknowledges the support of the research fellowships of the Japan
Society for the Promotion of Science for Young Scientists (17-10511).
The numerical simulations reported here were carried out on 
Fujitsu-made vector parallel processors VPP5000
kindly made available by the Astronomical Data Analysis
Center (ADAC) at National Astronomical Observatory of Japan (NAOJ)
for our  research project why36b.

\appendix

\section{Dependences of $M_{\rm h}-Z_{\rm gc}$ relations 
on $z_{\rm trun}$.}

Fig. A1 shows that the initially steep  $M_{\rm h}-Z_{\rm gc}$ relation
at $z=6$ becomes  flatter
in the model with $z_{\rm trun}=6$, though the degree of the flattening 
is less significant in this model than in the model with
$z_{\rm trun}=10$.
This reflects the fact that galaxies experienced a smaller number
of galaxy merging (which can flatten the  $M_{\rm h}-Z_{\rm gc}$ relation)
between $z=z_{\rm trun}$ and $z=0$ in this model.
Fig. A2 shows that the   $M_{\rm h}-Z_{\rm gc}$ relation is almost
flat in the model with $z_{\rm trun}=15$, which means  that
hierarchical merging of low-mass galaxies between $z_{\rm trun}$
and $z=0$ almost completely erases out the original steep 
 $M_{\rm h}-Z_{\rm gc}$ relation: this simulated
$M_{\rm h}-Z_{\rm gc}$ at $z=0$ is inconsistent with
the observed one.
These results shown in Figs. A1 and A2 thus suggest
that the degree of the flattening in the $M_{\rm h}-Z_{\rm gc}$ relation
depends on $z_{\rm trun}$ in the sense that models
with earlier truncation of GC formation
show the larger degree of the flattening.
The absence of low-mass galaxies with GCs 
for ${\log}_{10} {\rm M}_{\rm h} > 11.4$ (${\rm M}_{\odot}$)
in the model with  $z_{\rm trun}=15$
is inconsistent with observations.

\begin{figure}
\psfig{file=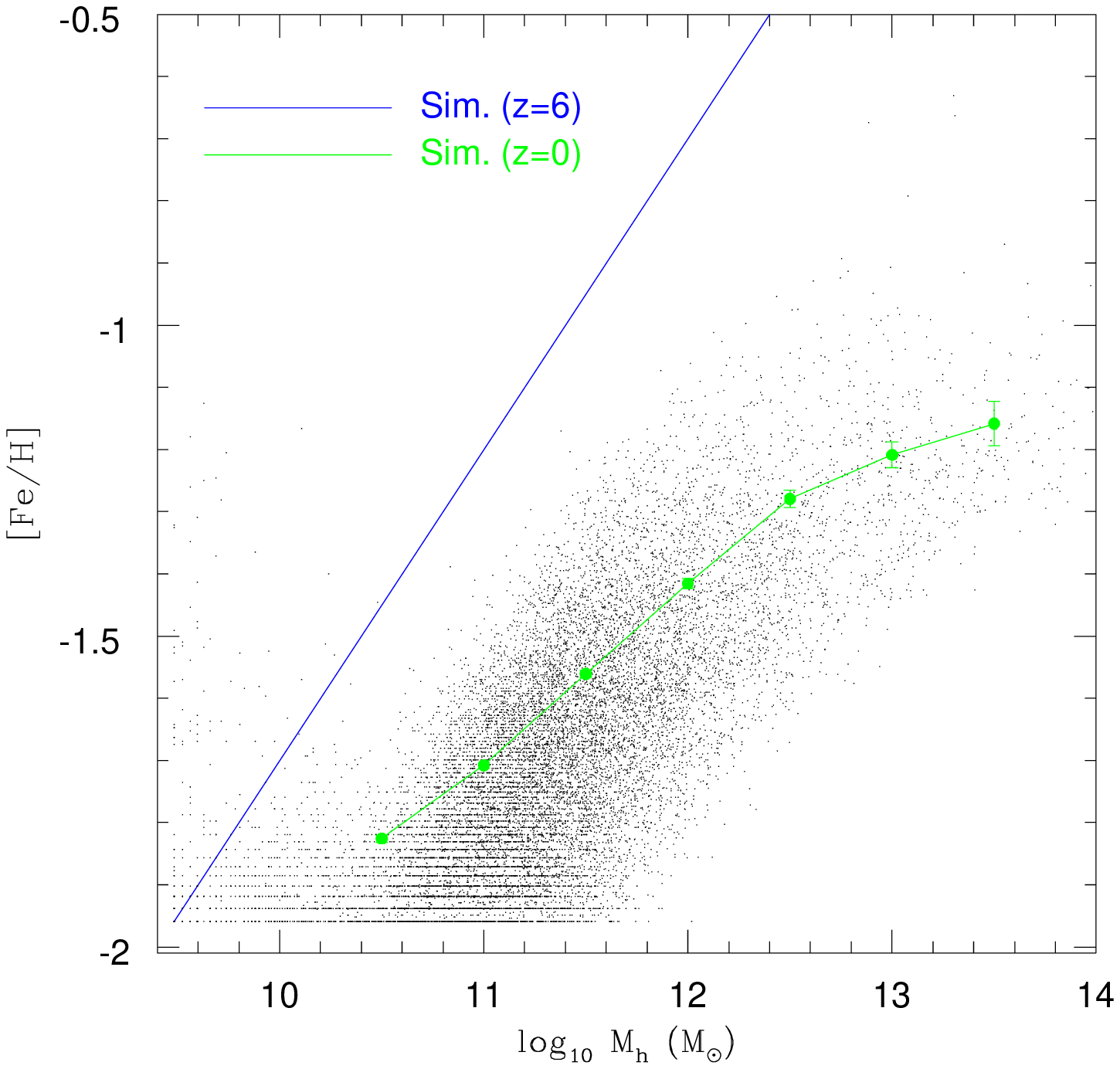,width=8.5cm}
\caption{ 
The same as Figure 2 but for the model with $z_{\rm trun}=6$.
}
\label{Figure. 14}
\end{figure}

\begin{figure}
\psfig{file=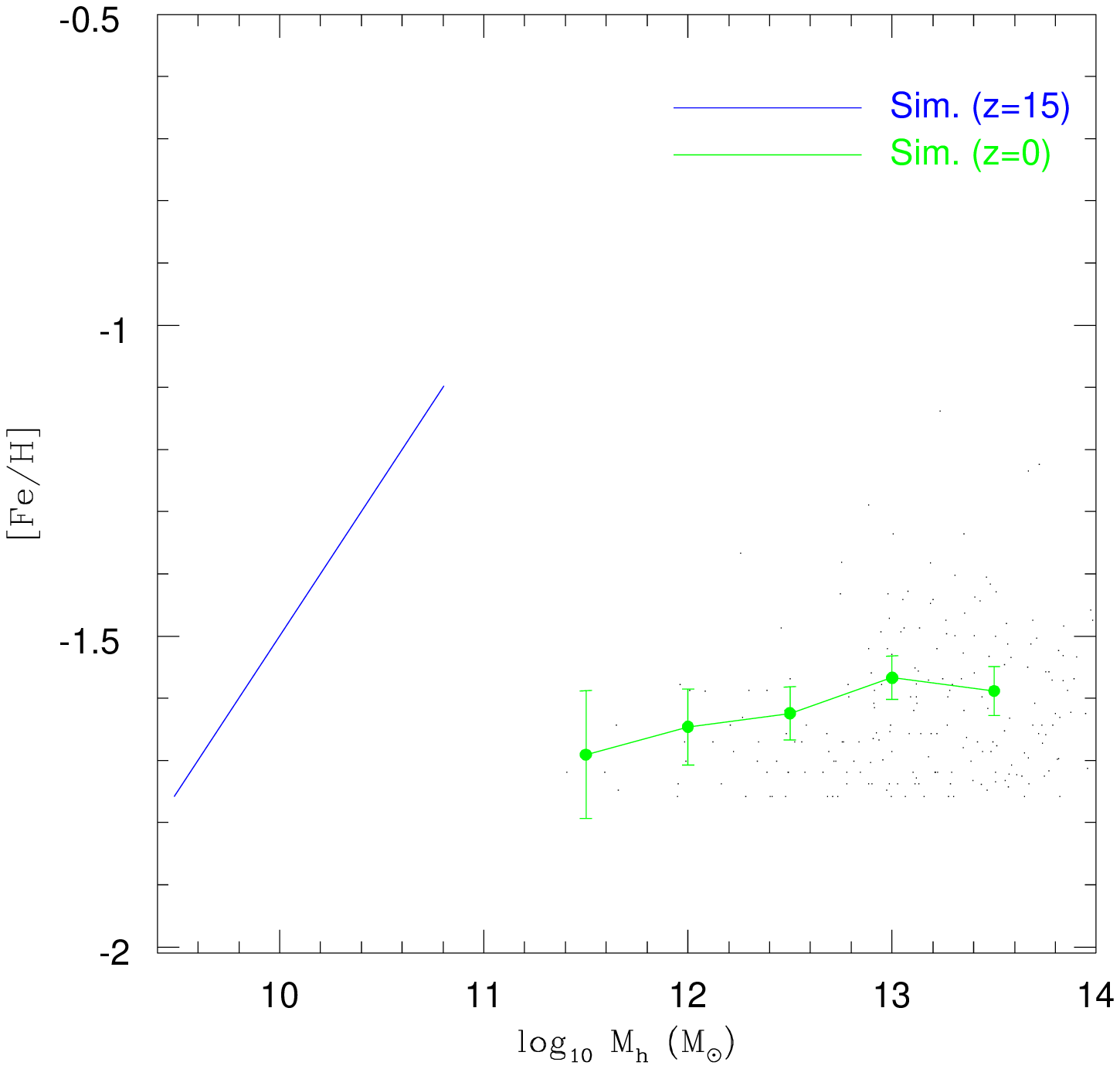,width=8.5cm}
\caption{ 
The same as Figure 2 but for the model with $z_{\rm trun}=15$.
}
\label{Figure. 15}
\end{figure}

\section{Dependences of $M_{\rm s}-Z_{\rm gc}$ relations 
on $M_{\rm s}/L$.}

Fig. B1 shows that the original $L-Z_{\rm gc}$ relation
becomes significantly flatter between $z_{\rm trun}$ and $z=0$
in the model with  $M_{\rm s}/L=5$.
The locations of galaxies on the $M_{\rm s}$-$Z_{\rm gc}$ plane
are shifted to the righter directions owing to the 
larger stellar masses of galaxies
in this model with higher $M_{\rm s}/L$. 
As a result of this,
the simulated $Z_{\rm gc}$ for a given $M_{\rm s}$ 
is slightly smaller than the observed one.
The change in $M_{\rm s}/L$ however does  not change
the final slope in the $L-Z_{\rm gc}$ relation
in comparison with the model with $M_{\rm s}/L=1$.
The flattening of the $L-Z_{\rm gc}$ 
relation between  $z_{\rm trun}$ and $z=0$
does not depend on $M_{\rm s}/L$,
as long as we adopt assumptions of (i)  a constant $M_{\rm s}/L$
at $z=z_{\rm trun}$ and  (ii) an initial
$L-Z_{\rm gc}$ relation of $Z_{\rm gc} \propto L^{0.5}$.
This means  that if we adopt a slightly higher value
of ${\beta}_{\rm gc}$ for the higher $M_{\rm s}/L$ model,
the observed relation can be better reproduced
under the above assumptions.
Thus  high values of ${\beta}_{\rm gc}$ are required in
the models with higher $M_{\rm s}/L$ 
and with the above assumptions
for the observation
to be well reproduced.

\begin{figure}
\psfig{file=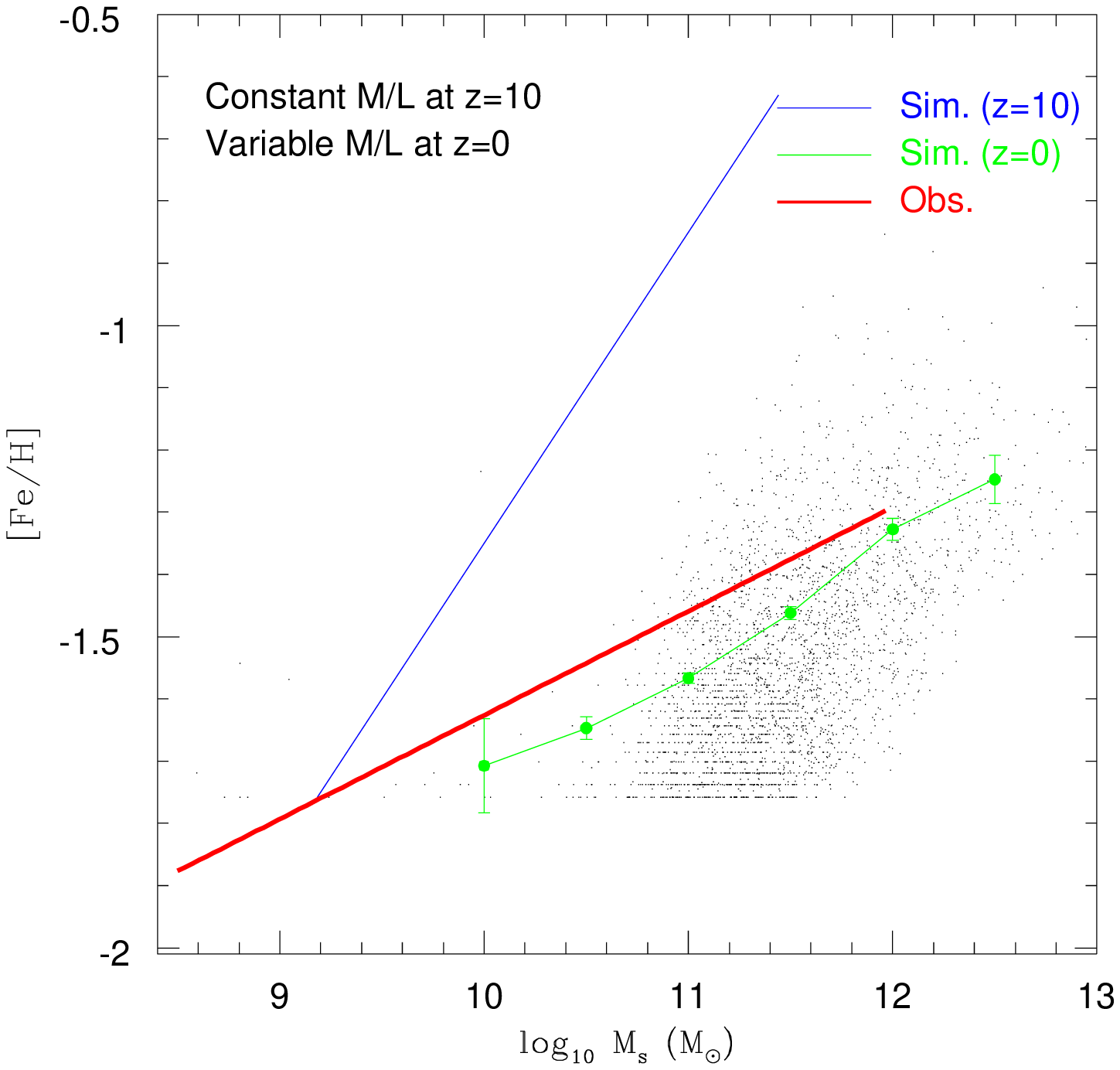,width=8.5cm}
\caption{ 
The same as Figure 3 but for the model with $M_{\rm s}/L=5$.
}
\label{Figure. 16}
\end{figure}

\section{Dependences of $M_{\rm s}-Z_{\rm gc}$ relations 
on $z_{\rm trun}$.}

Figs. C1 and C2 show the simulated $M_{\rm s}-Z_{\rm gc}$ relations
(thus $L-Z_{\rm gc}$ relations for the fixed $M_{\rm s}/L$ ratios)
for the models with $z_{\rm trun}$ = 6 and 15, respectively.
It is clear from these figures and Fig. 5 that
the model with $z_{\rm trun}=10$ can better reproduce
the observed  flat $L-Z_{\rm gc}$ relation (or $M_{\rm s}-Z_{\rm gc}$ one).
The model with $z_{\rm trun}=15$ shows the 
$L-Z_{\rm gc}$ relation flatter than the observed one
whereas  the model with $z_{\rm trun}=6$ shows 
the steeper one.
In the model with  $z_{\rm trun}=15$ ($z_{\rm trun}=6$),
a larger (smaller) number of merger events for a longer (shorter)
time scale between $z=z_{\rm trun}$ and $z=0$ 
can flatten the original $L-Z_{\rm gc}$ relation at $z=z_{\rm trun}$
to a larger (smaller)  extent.
This is the essential reason why 
the simulated $L-Z_{\rm gc}$ relation depends on
$z_{\rm trun}$.

Thus the growth of galaxies via hierarchical merging/accretion
of low-mass halos with and without MPCs between
$z=z_{\rm trun}$ and $z=0$ can flatten the original
$L-Z_{\rm gc}$ at  $z=z_{\rm trun}$
owing to merging/accretion of MPCs with different metallicities.
If $z_{\rm trun}$ corresponds to the epoch of reionization
($z_{\rm reion}$) and it is determined from future
observations,
then the results of numerical simulations shown in Figs. 6, 
C1, and C2 imply that 
we can infer the original $L-Z_{\rm gc}$ at  $z=z_{\rm reion}$
by comparing the simulations with the observed  $L-Z_{\rm gc}$
at $z=0$.

\begin{figure}
\psfig{file=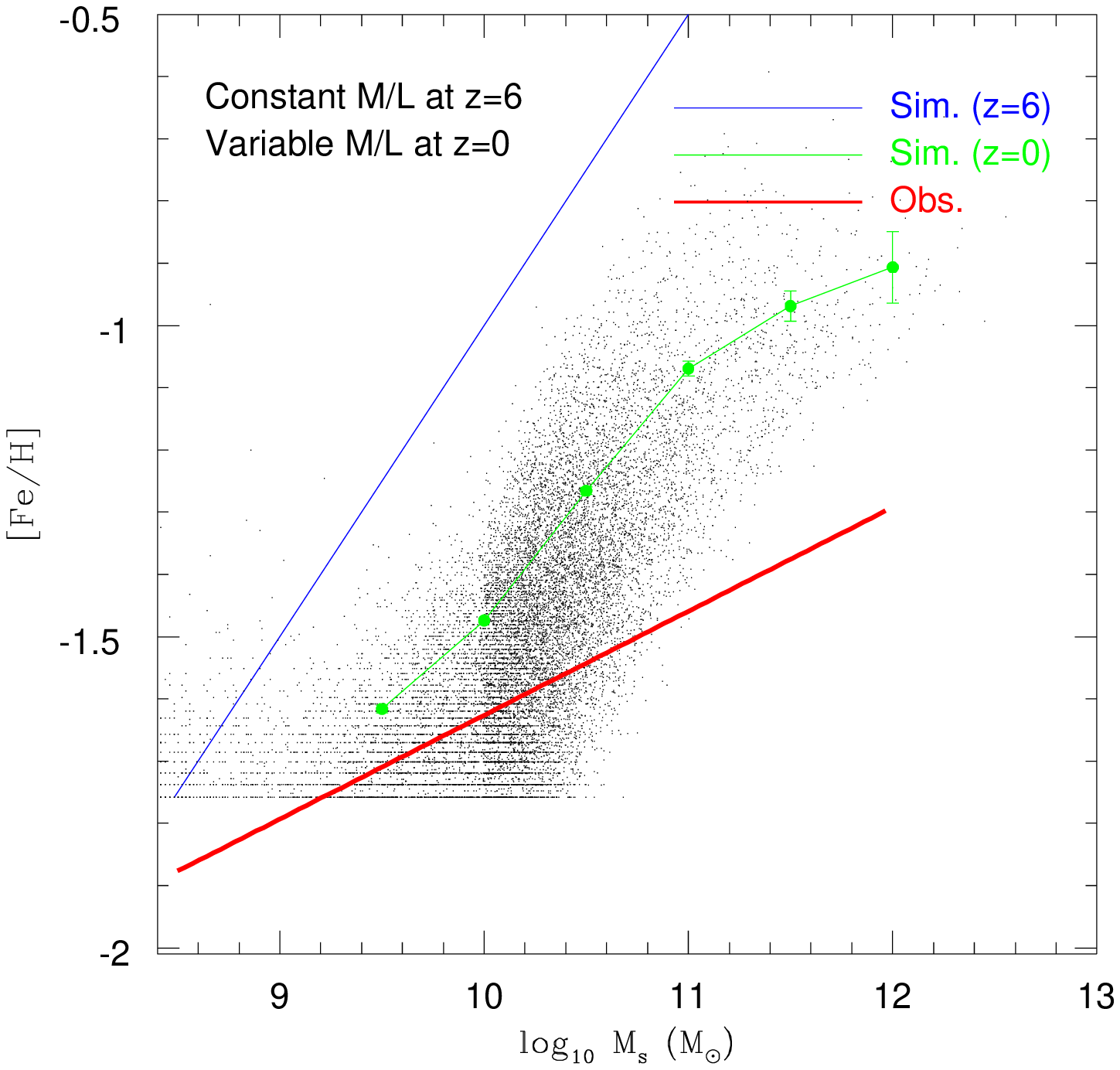,width=8.5cm}
\caption{ 
The same as Figure 3 but for the model with $z_{\rm trun}=6$.
}
\label{Figure. 17}
\end{figure}

\begin{figure}
\psfig{file=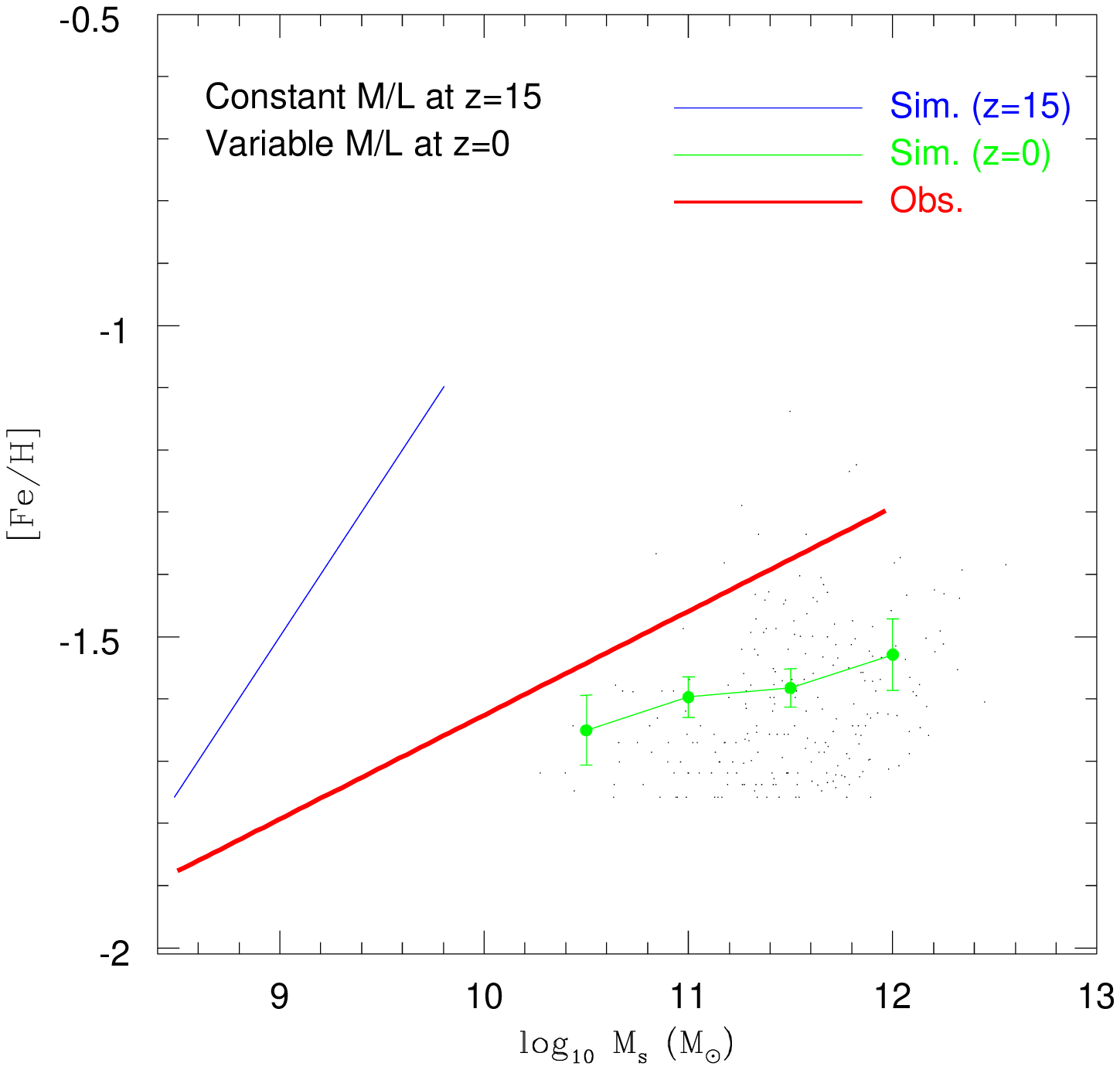,width=8.cm}
\caption{
The same as Figure 3 but for the model with $z_{\rm trun}=15$.
}
\label{Figure. 18}
\end{figure}

\section{Dependences of scaling  relations 
on $z_{\rm trun}$.}

Fig. D1 shows that  the dynamical correlations are  
different between different $z_{\rm trun}$, in particular, 
for $M_{\rm s}$-dependences of $I_{\rm e}$.
the   $M_{\rm s}$-dependence of $R_{\rm e}$ 
is steeper for higher $z_{\rm trun}$ whereas
the   $M_{\rm s}$-dependence of $\sigma$
is not so different between different $z_{\rm trun}$.
The slope in the  $M_{\rm s}$-dependence of $I_{\rm e}$
is negative for $z_{\rm trun}=10$ and 15 and positive
for $z_{\rm trun}=6$.  
The derived differences in the $M_{\rm s}$-dependences of
dynamical properties of GCSs between different $z_{\rm trun}$
suggest that future observational studies on the
$M_{\rm s}$-dependences can provide some constraints
on $z_{\rm trun}$.

Structural properties of GCSs
have been suggested to 
provide constraints on $z_{\rm trun}$
(Santos 2003; Bekki 2005; Moore et al. 2006). 
The present study suggested that if
(i) the adopted $L-Z_{\rm gc}$ relation 
at $z=z_{\rm trun}$ ($Z_{\rm gc} \propto L^{0.5}$)
is reasonable and realistic
and (ii) $M/L$ does not depend on galaxy masses at
$z=z_{\rm trun}$,
the observed $L-Z_{\rm gc}$ relation at $z=0$
could also provide  constraints on $z_{\rm trun}$. 
It is therefore an important observational test whether 
$z_{\rm trun}$ derived from
the observed $M_{\rm s}$-dependences 
is consistent with $z_{\rm trun}$ derived from
the structural and chemical properties of GCSs. 
%The observational data sets for dynamical properties
%of GCSs are currently unavailable so that we can not discuss
%the best possible $z_{\rm trun}$   derived from
%the observed $M_{\rm s}$-dependences.

\begin{figure}
\psfig{file=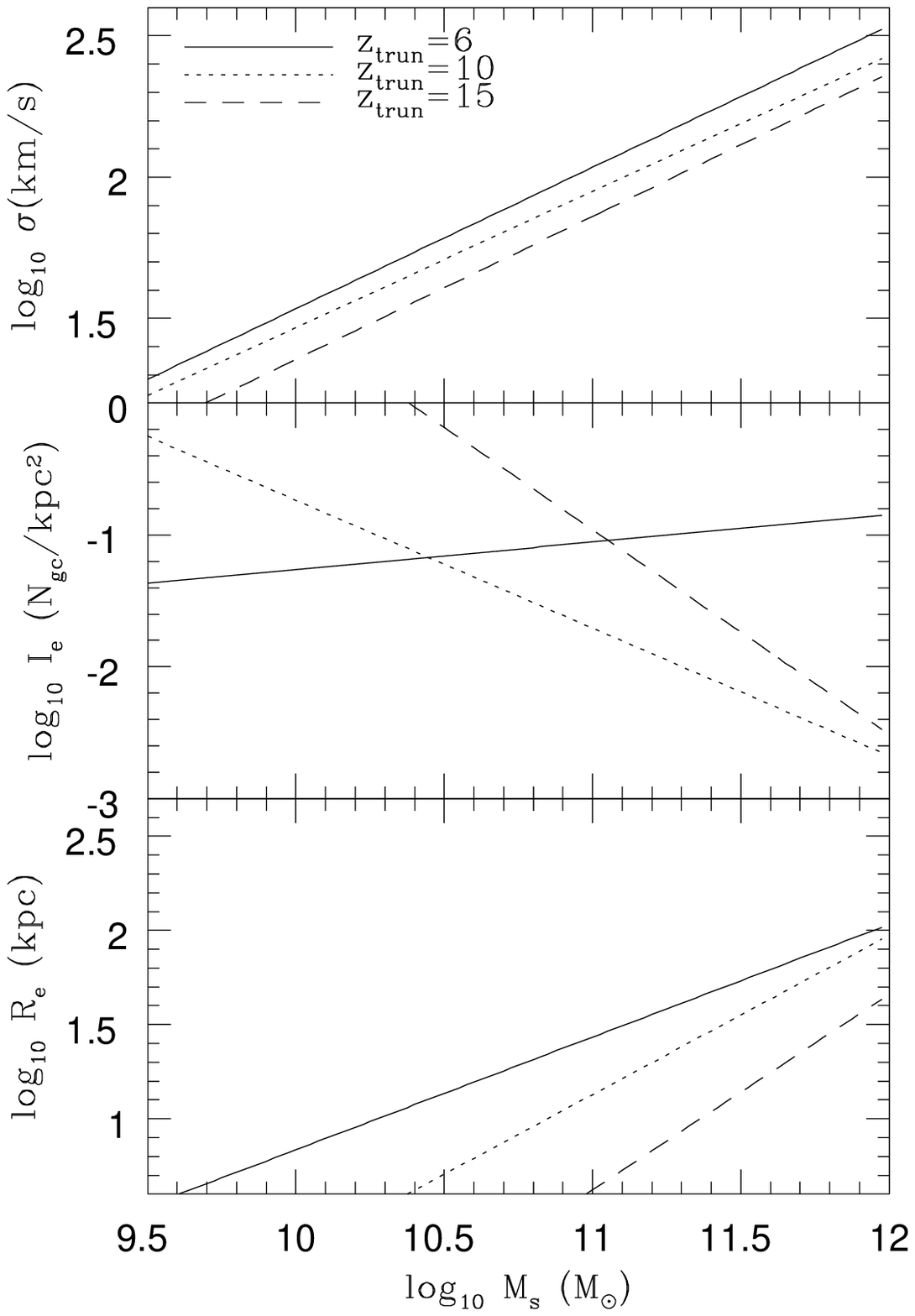,width=8.5cm}
\caption{ 
Dependences of line-of-sight  velocity dispersions ($\sigma$, top),
effective surface number densities ($I_{\rm e}$, middle),
and half-number radii ($R_{\rm e}$, bottom) of MPCs  on
$M_{\rm s}$  at $z=0$ for the model
with $z_{\rm trun}=6$ (solid),
$z_{\rm trun}=10$ (dotted),
and $z_{\rm trun}=15$ (dashed).
Each of these lines represent the least-square fit to
the simulation data for each of the three $M_{\rm s}$-dependences
in each model. 
}
\label{Figure. 20}
\end{figure}

\section{The influences of maximum GC masses on the blue tilt} 

We adopted an assumption of 
a universal  GC mass (luminosity) function
with the same lower ($m_{low}$) and upper 
(or luminosity) mass cut off ($m_{\rm upp}$)
for all halos in the present study.
Since recent observations have suggested a possible
maximum mass $m_{\rm max}$ of star clusters in galaxies
(e.g., Gieles et al. 2006)
and correlations between  $m_{\rm max}$ and physical properties
of their host galaxies, such as star formation rates
(e.g., Larsen \& Richtler 2000),
the above assumption on the fixed $m_{\rm upp}$
could be less realistic. 
We thus investigated the models with $m_{\rm max}-M_{\rm h}$
relations (i.e., $m_{\rm upp}-M_{\rm h}$)
suggested by numerical simulations of Kravtsov \& Gnedin 2005.

Fig. E1 shows the result of 
the model  with $M_{\rm h}=3.0 \times 10^{13} {\rm M}_{\odot}$
and $N_{\rm gn}=62$  for 
$m_{\rm max}=2.9 \times 10^6 {\rm M}_{\odot}
 {(\frac{M_{\rm h}}{10^{11} {\rm M}_{\odot}})}^{1.29}$.
Owing to the introduction of the $m_{\rm max}-M_{\rm h}$ relation,
the overall distribution  of MPCs  on the  $M_{\rm V}$-[Fe/H]
in this model appears to be only slightly more similar 
to the observed
blue tilt than the model shown in Fig. 11.
The least square fits to the simulation data show
that $M_{\rm V}  = -10.74 -1.62 \times {\rm [Fe/H]}$
(i.e., $Z \propto L^{1.55}$).
This $Z \propto L^{1.55}$ relation is however still
significantly steeper than the observed one
of $Z \propto L^{0.55}$ (Harris et al. 2006).
Although the models with steeper $m_{\rm max}-M_{\rm h}$ relations can
show flatter $Z-L$ relations,
the simulated dispersions in $M_{\rm V}$ for given
metallicity bins (in particular,  for [Fe/H] $>-1.5$)
are too large to be consistent with observations.
These  suggest that the models
with  $m_{\rm max}-M_{\rm h}$ relations  still miss some important
ingredients of GC formation.
We discuss these problems in our forthcoming papers (Bekki et al. 2007,
in preparation).

\begin{figure}
\psfig{file=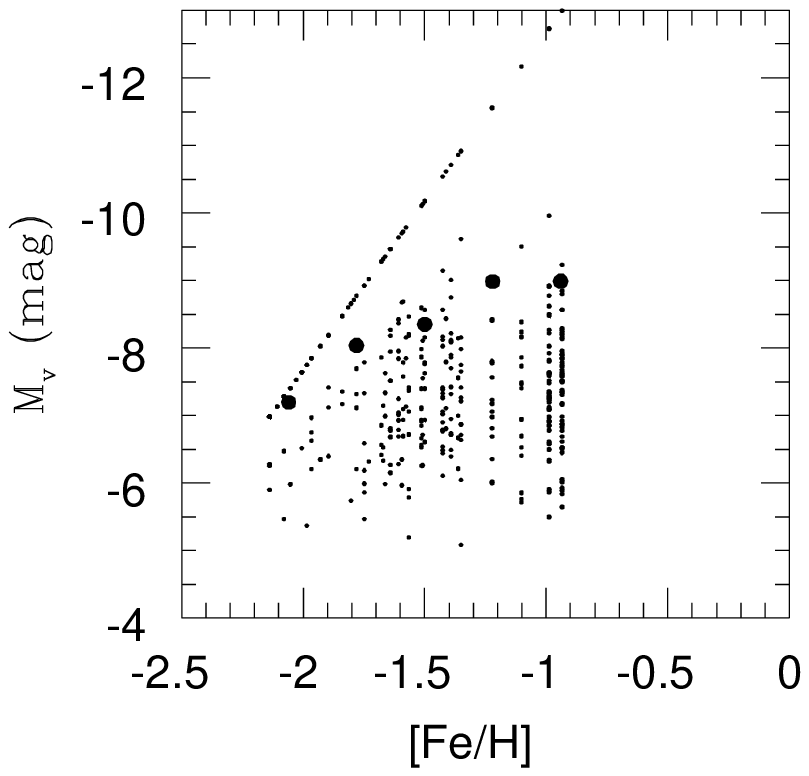,width=8.5cm}
\caption{ 
The same as Fig. 11 but for the model in which maximum GC masses
($m_{\rm max}$) in halos are assumed to correlate with
their host halo masses. 
}
\label{Figure. 21}
\end{figure}


\begin{thebibliography}{}

\bibitem[]{} 
Ashman, K. M.,  Zepf, S. E., 1992, ApJ, 384, 50

\bibitem[]{} 
Ashman, K. M.,  Zepf, S. E., 1998, in Globular cluster systems,
Cambridge, U. K. ; New York : Cambridge University Press,


\bibitem[]{} 
Bassino, L. P., Muzzio, J. C., \& Rabolli, M. 1994, ApJ, 431, 634

\bibitem[]{}
Baumgardt, H. 1988, A\&A, 330, 480
                                                              

%\bibitem[]{}  Bassino, L. P., Cellone, S. A., Forte, J. C., \&  Dirsch, B.
%2003, A\&A, 399, 489

\bibitem[]{} Beasley, M.~A., Baugh, C.~M., Forbes, D.~A., Sharples, R.~M.,
Frenk, C.~S.\ 2002, MNRAS, 333, 383

\bibitem[]{}
Beasley, M. A., Forbes, D. A.,
Brodie, J. P.,  Kissler-Patig, M., 2004, MNRAS, 347, 1150

\bibitem[]{} 
Bekki, K., 1998, ApJ, 496, 713

\bibitem[]{} 
Bekki, K.; Freeman, K. C., 2003, MNRAS, 346L, 11

\bibitem[]{} 
Bekki, K., Chiba, M., 2004, A\&A, 417, 437

\bibitem[]{} Bekki,~K. 2005, ApJ, 626, L93

\bibitem[]{} Bekki,~K., \& Chiba,~M. 2005, ApJ, 625, L107
 
%\bibitem[]{} 
%Bekki, K., Forbes, D. A.,  Beasley, M. A., \&  Couch, W. J.
%2002, MNRAS, 344, 1334

\bibitem[]{} 
Bekki, K.,  Beasley, M. A., Brodie, J. P., \& Forbes, D. A.
2005,  MNRAS, 363, 1211

\bibitem[]{} Bekki,~K.,  Forbes, D. A., 2006, A\&A, 445, 485

\bibitem[]{} 
Bekki, K., Yahagi, H., Forbes, D. A., 2006, ApJL, 645, 29

\bibitem[]{} 
Bekki, K., \& Yahagi, H. 2006, accepted in MNRAS 

\bibitem[]{} Bertschinger,~E. 1995, astro-ph/9506070 

\bibitem[]{} Bertschinger,~E. 2001, ApJS, 137, 1

\bibitem[]{} 
Binney, J.,  Tremaine, S., 1987 in Galactic Dynamics.

\bibitem[]{} 
Brodie, J. P., Strader, J., 2006, ARA\&A in press (astro-ph/0602601)

\bibitem[]{} 
Bromm, V.,  Clarke, C. J., 2002, ApJL, 566, 1

\bibitem[]{} 
Capelato, H. V., de Carvalho, R. R., 
Carlberg, R. G.,  1995, ApJ, 451, 525


\bibitem[]{} 
Capuzzo-Dolcetta, R., 2006, in the Globular Clusters to Guides to
Galaxies (astro-ph/0605162)


\bibitem[]{} 
Cohen, J. G., 2000, AJ, 119, 162

\bibitem[]{} 
Cole, S., Lacey, C., Baugh, C., Frenk, C., 2000, MNRAS, 319, 168

\bibitem[]{} 
C\^ote,~P.,
Marzke, R. O., West, M. J.  1998, ApJ, 501, 554 

\bibitem[]{}
C\^ote,~P. et al. 2001, ApJ, 559, 828


\bibitem[]{}
C\^ote,~P. et al., 2006, accepted in ApJS (astro-ph/0603252)

\bibitem[]{}
C\^ote,~P.  McLaughlin, D. E.,
Cohen, J. G.,  Blakeslee, J. P., 2002, ApJ, 591, 850


\bibitem[]{} 
Dantas, C. C., Capelato, H. V., Ribeiro, A. L. B., 
de Carvalho, R. R. 2003, MNRAS, 340, 398



\bibitem[]{} 
Davis, M., Efstathiou, G., Frenk, C. S., \&  White, S. D. M.
1985, ApJ, 292, 371

\bibitem[]{} 
Dekel, A.,  Silk, J., 1986, ApJ, 303, 39

\bibitem[]{} 
Dekel, A., Woo, J., 2003, MNRAS, 344, 1131
 

\bibitem[]{} 
Djorgovski, S.,  Davis, M. 1987, ApJ, 313, 59


\bibitem[]{} 
Fan, X. et al. 2003, AJ, 125, 1649 


\bibitem[]{} 
Forbes, D. A., Forte, J. C., 2001, MNRAS, 322, 257

\bibitem[]{} 
Forbes, D. A., 2005, ApJL, 635, 137 

%\bibitem[]{} 
%Forte, J. C.; Martinez, R. E.,   Muzzio, J. C., 1982, AJ, 87, 1465

\bibitem[]{} 
Freeman, K. C. 1993, in The globular clusters-galaxy connection,
edited by Graeme H. Smith, and Jean P. Brodie,
ASP conf. ser. 48, p608

\bibitem[]{} 
Gieles, M., Larsen, S. S., Bastian, N., Stein, I. T.,
2006, A\&A, 450, 129

%\bibitem[]{} 
%Jord\'an, A.,  West, M. J.,
%C\^ote,~P., \&  Marzke, R. O. 2003, AJ, 125, 1642 

\bibitem[]{} 
Harris, W. E.,  1991, ARA\&A, 29, 543 

\bibitem[]{} 
Harris, W. E., Whitmore, B. C.,
Karakla, D.,  Oko\'n, W., 
Baum, W. A., Hanes, D. A.,  Kavelaars, J. J.,
2006, ApJ, 636, 90

\bibitem[]{} 
Kissler-Patig, M., Gebhardt, K.,  1998, AJ, 116, 2237 

\bibitem[]{} Kogut, A. et al. 2003, ApJs, 148, 161

\bibitem[]{} 
Kormendy, J., Freeman, K. C., 2004, 
IAU Symposium no. 220,  
Edited by  S. D. Ryder, D. J. Pisano, M. A. Walker, and K. C. Freeman. 
San Francisco: Astronomical Society of the Pacific., p.377

\bibitem[]{} 
Kravtsov, A. V., Gnedin, O. Y., 2005, ApJ, 623, 650

\bibitem[]{} 
Larsen, S. S.,  Richtler, T., 2001, A\&A. 354, 836

\bibitem[]{} 
Larsen, S. S., Brodie, J. P., Huchra, J. P.,
Forbes, D. A., Grillmair, C. J., 2001, AJ, 121, 2974

\bibitem[]{} 
Lotz, J. M., Miller, B. W., Ferguson, H. C., 2004, ApJ, 613, 262

\bibitem[]{} 
Marinoni, C., Hudson, M. J., 2002, ApJ, 569, 101


\bibitem[]{} 
Mashchenko, S.,  Sills, A., 2005, 619, 258

%\bibitem[]{} Mieske, S., Hilker, M., \&  Infante, L.	
%2004, A\&A, 418, 445
	
\bibitem[]{} 
Mieske, S., Hilker, M.,   Infante, L.,	 Jord\'an, A.,
2006, AJ, 131, 2442

\bibitem[]{} 
Mieske, S., et al. 2006, accepted by ApJ (astro-ph/0609079) 

\bibitem[]{} 
Minniti, D., Meylan, G., Kissler-Patig, M., 1996, A\&A, 312, 49


\bibitem[]{} 
Moore, B., et al. 2006, astro-ph/0510370

\bibitem[]{} 
Navarro, J. F., Frenk, C. S., \& White, S. D. M. 1996, ApJ, 462, 563 (NFW)

\bibitem[]{} 
Page, L. et al., 2006, submitted to ApJ (astro-ph/0603450)

\bibitem[]{} 
Peng, E. W., Ford, H. C., Freeman, K. C., 2004, ApJ, 602, 705

\bibitem[]{} 
Peng, E., et al. 2006, ApJ, 639, 95


\bibitem[]{} 
Pierce, M. et al, 2006, MNRAS, 366, 1253

\bibitem[]{} 
Rhode, K. L., Zepf, S. E.,  \& Santos, M. R. 2005, ApJL, 630, 21

\bibitem[]{} 
Rhode, K. L., Zepf, S. E.,  2004, AJ, 127, 302 

\bibitem[]{} 
Richtler, T., 2003,  Stellar Candles for the Extragalactic Distance Scale,
Edited by D. Alloin and W. Gieren, vol. 635, p.281

\bibitem[]{} 
Richtler, T. et al. 2004, AJ, 127, 2094

\bibitem[]{} 
Romanowsky, A. J., 2006, in the Globular Clusters to Guides to
Galaxies.

\bibitem[]{} 
Santos, M. R. 2003, 
in Extragalactic Globular Cluster Systems, Proceedings 
of the ESO Workshop,  p. 348

\bibitem[]{} 
Spitler, L. R., Larsen, S. S., 
Strader, J., Brodie, J. P., Forbes, D. A.,  Beasley, M. A.,
2006, accepted in AJ (astro-ph/0606337)

\bibitem[]{} 
Strader, J.,  Brodie, J. P., Forbes, D. A., 2004, AJ, 127, 3431
 
\bibitem[]{} 
Strader, J.,  Brodie, J.  P., Spitler, L., Beasley, M. A.,
2006, submitted to AJ (astro-ph/0508001).

\bibitem[]{} 
Susa, H.,  Umemura, M. 2004, ApJ, 600, 1

%\bibitem[]{} van~den~Bergh,~S. 1958, The Observatory, Vol. 78, p85. 

\bibitem[]{} 
van~den~Bergh,~S. 1986, AJ, 91, 271


\bibitem[]{} van~den~Bergh,~S. 2000,  
The Galaxies of the Local Group, Cambridge: Cambridge Univ. Press.

\bibitem[]{} 
Vazdekis, A., Casuso, E., Peletier, R. F., 
Beckman, J. E., 1996, ApJS, 106, 307

\bibitem[]{} 
van de Ven, G., van Dokkum, P. G.,  Franx, M., 2003, MNRAS, 344, 924


\bibitem[]{} 
Walcher, J. et al., 2006, submitted to ApJ (astro-ph/0604140)


\bibitem[]{} 
West, M. J.,
C\^ote,~P.,
Marzke, R. O.,   Jord\'an, A.,  2004, Nat, 427, 31




%\bibitem[]{} 
%Yahagi, H. 2005, Doctoral Thesis, University of Tokyo

\bibitem[]{} 
Yahagi, H. 2005, PASJ,  57, 779

\bibitem[]{} 
Yahagi, H., \&  Yoshii, Y. 2001, ApJ, 558, 463

\bibitem[]{} 
Yahagi, H., Nagashima, M., \& Yoshii, Y., 2004, ApJ, 605, 709

\bibitem[]{} 
Yahagi, H., Bekki, K., 2005, MNRAS, 364L, 86

\bibitem[]{} 
Zaritsky, D., Gonzalez, A., Zabludoff, A., 2006, ApJ, 642, 37

 
\bibitem[]{} 
Vesperini, E., Zepf, S. E., Kundu, A.,  Ashman, K. M., 2003, ApJ, 593, 760


\bibitem[]{} 
Zepf, S. E., Beasley, M. A., Bridges, T. J.,
Hanes, D. A., Sharples, R. M., Ashman, K. M.,  Geisler, D.,
2000, AJ, 120, 2928

\bibitem[]{} 
Zinnecker, H., Keable, C. J., Dunlop, J. S., Cannon, R. D.,   Griffiths,  W. K.,
1988, in Grindlay, J. E., Davis Philip A. G., eds, Globular cluster systems in Galaxies,
Dordrecht, Kluwer, p603 


\end{thebibliography}
\end{document}